\documentclass[10pt,journal]{IEEEtran}

\usepackage{amssymb}
\usepackage{amsmath}
\usepackage{cite}
\usepackage{url}
\usepackage{xcolor}
\usepackage{cite,graphicx,amsmath,amssymb}
\usepackage{fancyhdr}
\usepackage{mdwmath}
\usepackage{mdwtab}
\usepackage{caption}
\usepackage{amsthm}
\usepackage{setspace}
\usepackage{hyperref}
\usepackage{algorithm}
\usepackage{algorithmic}
\usepackage{multirow}
\usepackage{makecell}
\usepackage{mathtools}
\usepackage{subcaption}
\usepackage{bm}
\usepackage{tikz}
\usepackage{xurl}

\hypersetup{colorlinks=true,
linkcolor=blue,
citecolor=blue,      
urlcolor=black,
}

\graphicspath{{./src/img/}}

\newtheorem{remark}{Remark}
\newtheorem{theorem}{Theorem}

\newtheorem{lemma}{Lemma}

\newtheorem{corollary}{Corollary}

\allowdisplaybreaks
\NewDocumentCommand{\multiubrace}{mmm}
 {
  \egreg_multiubrace:nnn {#1} {#2} {#3}
 }

\title{Modeling and Beamforming Optimization for Pinching-Antenna Systems}

\author{
        Zhaolin Wang,~\IEEEmembership{Member,~IEEE,}
        Chongjun Ouyang,~\IEEEmembership{Member,~IEEE,} Xidong Mu,~\IEEEmembership{Member,~IEEE,}\\
        Yuanwei Liu,~\IEEEmembership{Fellow,~IEEE,} and Zhiguo Ding,~\IEEEmembership{Fellow,~IEEE,}
\thanks{Zhaolin Wang and Chongjun Ouyang are with the School of Electronic Engineering and Computer Science, Queen Mary University of London, London E1 4NS, U.K. (e-mail: \{zhaolin.wang, c.ouyang\}@qmul.ac.uk).}
\thanks{Xidong Mu is with the Centre for Wireless Innovation (CWI), Queen's University Belfast, Belfast, BT3 9DT, U.K. (e-mail: x.mu@qub.ac.uk).} 
\thanks{Yuanwei Liu is with the Department of Electrical and Electronic Engineering, The University of Hong Kong, Hong Kong (e-mail: yuanwei@hku.hk).}
\thanks{Zhiguo Ding is with Khalifa University, Abu Dhabi, UAE, and the University of Manchester, Manchester, M1 9BB, U.K. (e-mail: zhiguo.ding@manchester.ac.uk).}
\vspace{-0.3cm}
}

\begin{document}

\maketitle
\begin{abstract}
    The Pinching-Antenna SyStem (PASS) is a revolutionary flexible antenna technology designed to enhance wireless communication by establishing strong line-of-sight (LoS) links, reducing free-space path loss and enabling antenna array reconfigurability. PASS uses dielectric waveguides with low propagation loss for signal transmission, radiating via a passive pinching antenna, which is a small dielectric element applied to the waveguide. This paper first proposes a physics-based hardware model for PASS, where the pinching antenna is modeled as an open-ended directional coupler, and the electromagnetic field behavior is analyzed using coupled-mode theory. A simplified signal model characterizes the coupling effect between multiple antennas on the same waveguide. Based on this, two power models are proposed: equal power and proportional power models. Additionally, a transmit power minimization problem is formulated/studied for the joint optimization of transmit and pinching beamforming under both continuous and discrete pinching antenna activations. Two algorithms are proposed to solve this multimodal optimization problem: the penalty-based alternating optimization algorithm and a low-complexity zero-forcing (ZF)-based algorithm. Numerical results show that 1) the ZF-based low-complexity algorithm performs similarly to the penalty-based algorithm, 2) PASS reduces transmit power by over 95\% compared to conventional and massive MIMO, 3) discrete activation causes minimal performance loss but requires a dense antenna set to match continuous activation, and 4) the proportional power model yields performance comparable to the equal power model.
\end{abstract}


\begin{figure*}[t!]
    \centering
    \includegraphics[width=0.7\textwidth]{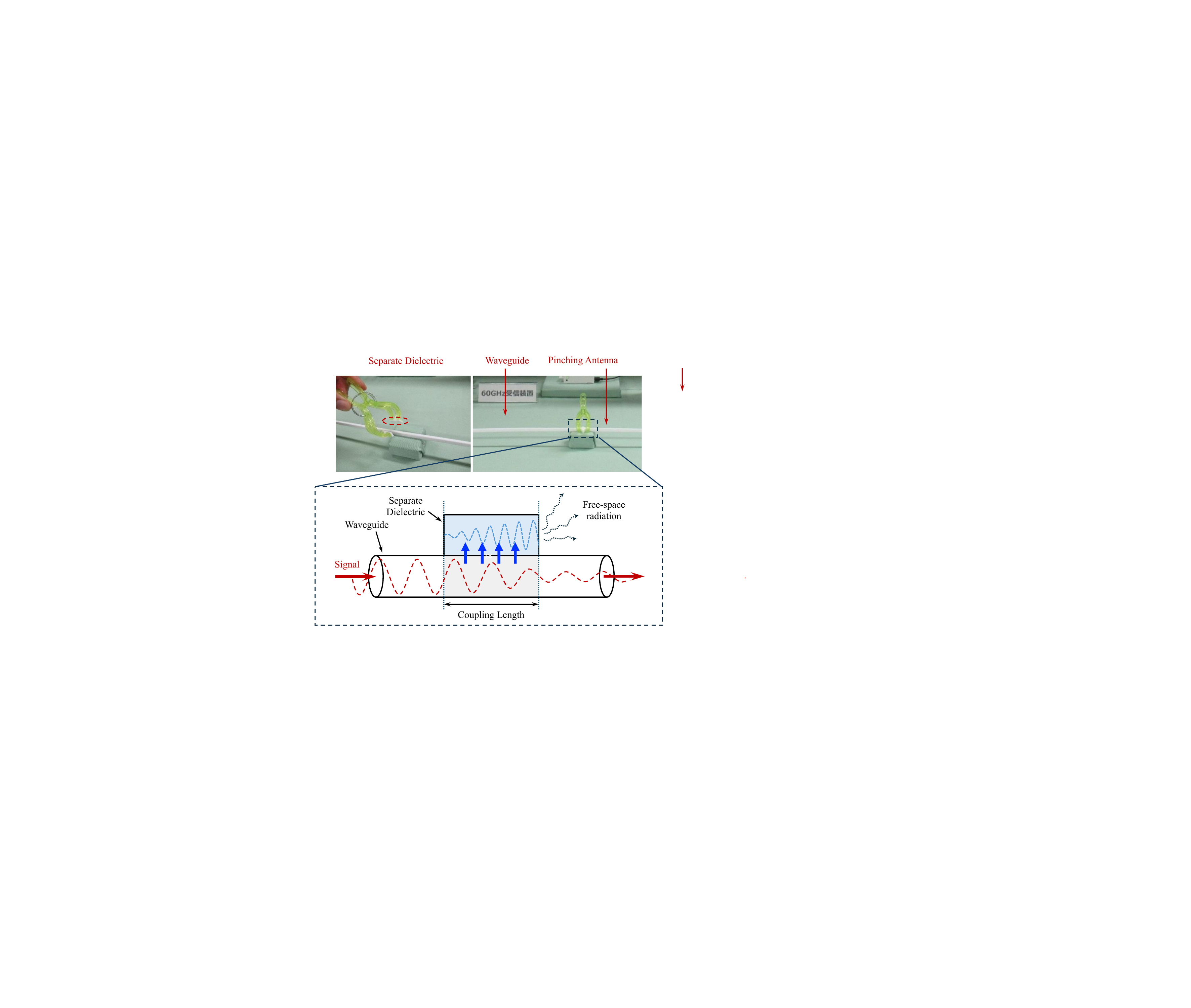}
    \caption{\textcolor{black}{Schematic illustration of the conventional wireless system (left) and the PASS (right).}}
    \label{comparison}
    \vspace{-0.5cm}
\end{figure*} 

\section{Introduction} \label{sec:intro}

\IEEEPARstart{S}INCE Marconi demonstrated the feasibility of wireless communication in the late 19th century, the technology has undergone significant evolution and remarkable transformations. To address the unpredictable and dynamic nature of wireless channels, numerous advancements have been made in the air interface design, channel coding, source compression, and communication protocols for improving data rates and enhancing reliability. Among these advancements, multiple-input multiple-output (MIMO) has been one of the most important evolutionary techniques for wireless communication over the past few decades. By exploiting antenna arrays, MIMO brings about multiple benefits, such as enhanced signal strength through beamforming, mitigation of multi-path fading, and efficient spatial-domain multiplexing of users~\cite{bjornson2023twenty}. Since the advent of the third generation (3G) system, MIMO has been a fundamental component of wireless communication standards. However, during that era, the size of antenna arrays in MIMO systems was generally limited. The breakthrough came when Marzetta demonstrated the significant benefits of deploying an infinite number of antennas in 2010~\cite{marzetta2010noncooperative}, revealing the potential of MIMO to enhance communication performance while reducing system complexity. This revelation paved the way for the concept of massive MIMO, i.e., employing large-scale antenna arrays at base stations. Over time, massive MIMO has evolved into a key research focus and has become a reality with the deployment of 5G networks.

However, massive MIMO has faced numerous challenges, as it is expected to transition from “Massive” in 5G (typically with 32-64 antennas) to “Gigantic” in 6G~\cite{Xtext, bjornson2024enabling}, where the number of antennas is expected to scale to hundreds or even thousands. One of the key obstacles is the complexity and cost of implementing massive MIMO since each antenna typically needs to be fed by a dedicated radio-frequency (RF) chain. Exploiting low-resolution analog-to-digital converters in RF chains or hybrid analog-digital antenna arrays with a limited number of RF chains were common methods to address this challenge, especially in the millimeter-wave band~\cite{heath2016overview}. More recently, advancements in metamaterials have paved the way for new antenna technologies, exemplified by waveguide-fed metasurface antennas~\cite{smith2017analysis, shlezinger2021dynamic, di2024reconfigurable}, which facilitate the ultra-dense deployment of antenna elements at a significantly lower cost and making massive MIMO implementation more feasible.

Flexible-antenna technique is a new evolution of MIMO. Unlike massive MIMO focusing on enlarging the wireless channel dimension, the flexible-antenna technique focuses on enabling the reconfiguration of the wireless channel. One of the most well-known approaches in this domain is the reconfigurable intelligent surface (RIS) technique~\cite{huang2019reconfigurable, wu2019intelligent, mu2021simultaneously}. By deploying RIS between transceivers, the wireless channel can be intelligently reconfigured by adjusting the phase shifts of the signals reflected/refracted by the RIS. More recently, fluid antennas~\cite{new2024tutorial} and movable antennas~\cite{zhu2023movable} have emerged as promising flexible-antenna technologies. The fundamental concept behind these approaches is to implement antenna arrays where individual antenna elements can dynamically adjust their positions within a spatial region, thus creating favorable channel conditions to enhance communication performance. 

Nevertheless, as shown on the left of Fig. \ref{comparison}, both massive MIMO and flexible-antenna techniques have limited capability in fundamentally addressing free-space pathloss and line-of-sight (LoS) blockage, two major causes of signal attenuation in wireless communications. While massive MIMO can achieve high beamforming gains to strengthen signals, it cannot combat LoS blockage and to effectively mitigate free-space pathloss, particularly for cell-edge users. RISs have been considered as a promising solution to overcome LoS blockage by creating virtual LoS paths. However, the double fading effect caused by signal reflection results in much higher pathloss compared to a direct LoS channel~\cite{ozdogan2019intelligent}. Additionally, fluid and movable antennas are typically capable of adjusting their positions only within a few wavelengths, making them more effective for mitigating small-scale fading rather than addressing large-scale pathloss. It is worthy to point out that all the aforementioned MIMO systems are lack of antenna array reconfigurability, i.e., once an antenna array is built, adding or removing antennas is no longer possible.

Pinching-Antenna SyStem (PASS) is a revolutionary technique for addressing the challenges of free-space pathloss and LoS blockage encountered by conventional multi-antenna technologies. This technique was originally proposed and prototyped by NTT DOCOMO in 2022~\cite{suzuki2022pinching}. As illustrated on the right of Fig. \ref{comparison}, PASS employs a dielectric waveguide as its primary transmission medium, which is known for its exceptionally low propagation loss (e.g., 0.01 dB/m \cite{yeh2000communication}). By pinching a small separated dielectric element, referred to as a \emph{pinching antenna}, onto the waveguide, the system enables signal emission from the waveguide into the pinching antenna, which then radiates the signal into free space. Building on this principle, waveguides can be pre-deployed to extend service coverage, allowing pinching antennas to be placed at positions close to users. This strategic placement transforms the wireless system into a \emph{near-wired} system and hence establishes strong LoS links with users, effectively minimizing free-space path loss and mitigating blockage issues, sharing a similar concept with surface wave communications where waves propagate over a slit along a metasurface and radiate from pre-defined locations \cite{9210135}. \textcolor{black}{These unique advantages make PASS a promising alternative for indoor communications}. Additionally, unlike existing MIMO systems and surface wave communications, PASS allows both the number and positions of pinching antennas (i.e., the radiation point) to be easily adjusted by simply pinching them to or releasing them from the waveguide~\cite{suzuki2022pinching}. This feature provides a low-cost and scalable approach to implementing MIMO while also facilitating the so-called \emph{pinching beamforming}, which enhances communication performance by dynamically optimizing antenna positions \cite{liu2025pinching}.

Given the successful prototyping of PASS by NTT DOCOMO, theoretical research on this topic has been steadily growing, though it remains in its early stages. The first theoretical study on PASS for the communication system design was presented in \cite{ding2024flexible}, where the authors provided a comprehensive analysis and developed low-complexity pinching beamforming designs for fundamental single-user and two-user scenarios. The array gain achieved by multiple pinching antennas on a waveguide was analyzed in \cite{ouyang2025array}, unveiling the optimal number of antennas and their spacing for maximizing the beamforming gain. The authors of \cite{bereyhi2025downlink} explored a downlink PASS architecture utilizing multiple waveguides, each equipped with a single pinching antenna, and proposed a greedy approach for jointly optimizing the transmit and pinching beamforming. Meanwhile, \cite{guo2025deep} examined a more generalized scenario, where multiple pinching antennas were deployed on each waveguide, and introduced a graph neural network-based deep learning method to address the corresponding joint beamforming optimization problem.

Although PASS has attracted growing attention, several key challenges remain unsolved. On the one hand, the physics modeling of PASS is still underdeveloped, which is crucial for establishing an accurate signal model. In existing studies \cite{ding2024flexible, ouyang2025array, bereyhi2025downlink, guo2025deep}, it is commonly assumed that all signal power within the waveguide is fully radiated into free space and that each pinching antenna on a waveguide emits identical radiation power—an assumption analogous to conventional MIMO systems. However, pinching antennas operate fundamentally differently from traditional electronic antennas, and such assumptions may lack a solid physical foundation and fail to accurately reflect real-world behaviors. On the other hand, most existing works design PASS under simplified assumptions \cite{ding2024flexible, ouyang2025array, bereyhi2025downlink}, such as a single user, a single waveguide, a single pinching antenna per waveguide, or perfectly aligned signal phases. Although the GNN-based deep learning model proposed in \cite{guo2025deep} is capable of handling more complex scenarios with arbitrary numbers of users, waveguides, and pinching antennas, it suffers from a key limitation: the model parameters need to be retrained once the system configuration changes, limiting its generalization ability. Motivated by these challenges, this paper aims to develop a fundamental physics-based signal model for PASS and explore joint beamforming designs for more general scenarios. The key contributions of this work are summarized as follows:
\begin{itemize}
    \item We propose a physics-based hardware model for PASS, in which a pinching antenna is modeled as an open-ended directional waveguide coupler to facilitate the adjustment of radiation characteristics and simplify signal modeling. Based on this model, we characterize the relationship between the electromagnetic (EM) fields within the waveguide and those radiated by the pinching antennas using coupled-mode theory.
    \item \textcolor{black}{We derive a novel signal model for PASS based on the proposed physics framework, assuming identical effective refractive indices for the waveguides and pinching antennas.} This model reveals the inherent coupling effect between the radiation power of pinching antennas deployed on the same waveguide. Leveraging this coupling relationship, we introduce two simplified power models and their respective implementation methods: equal power and proportional power models.
    \item We formulate a joint transmit and pinching beamforming optimization problem to minimize the transmit power in a general PASS system with arbitrary numbers of users, waveguides, and pinching antennas, considering both continuous and discrete activation of pinching antennas. To solve this highly nonconvex, coupled, and multimodal optimization problem, we propose two algorithms: the penalty-based alternating optimization algorithm and the zero-forcing (ZF)-based low-complexity algorithm.
    \item We provide comprehensive numerical results to validate the advantages of PASS and the effectiveness of the proposed algorithm. The results demonstrate that 1) the ZF-based algorithm delivers performance comparable to the penalty-based algorithm but has a low complexity, 2) PASS significantly reduces transmit power, achieving a reduction of over 95\% compared to conventional and massive MIMO, 3) a dense set of available antenna positions is required for discrete activation to achieve similar performance to continuous activation, and 4) the proportional power model exhibits performance comparable to the equal power model.
\end{itemize}

The rest of this paper is structured as follows. Section \ref{sec:model} introduces the proposed physics-based hardware model and signal model for PASS. Section \ref{sec:beamforing} presents the general system model for downlink PASS and introduces a penalty-based alternating optimization method and a ZF-based algorithm for solving the joint beamforming optimization problem. Numerical evaluations and performance comparisons under various system configurations are presented in Section \ref{sec:results}. Finally, Section \ref{sec:conclusion} summarizes the findings and concludes the paper.

\emph{Notations:} Scalars are denoted using regular typeface, vectors and matrices are represented in boldface, and Euclidean subspaces are indicated with calligraphic letters. The set of complex and real numbers are denoted by $\mathbb{C}$ and $\mathbb{R}$, respectively. The inverse, conjugate, transpose, conjugate transpose, and trace operators are denoted by $(\cdot)^{-1}$, $(\cdot)^*$, $(\cdot)^T$, $(\cdot)^H$, and $\mathrm{tr}(\cdot)$, respectively. The absolute value, Euclidean norm, Frobenius norm, and maximum norm are denoted by $|\cdot|$, $\|\cdot\|$, $\|\cdot\|_F$, and $\|\cdot\|_\infty$ respectively. The real part of a complex number of demoted by $\Re \{\cdot\}$. The entry in the $n$-th row and $m$-th column of a matrix $\mathbf{X}$ is denoted by $[\mathbf{X}]_{n,m}$. An identity matrix of dimension $N \times N$ is denoted by $\mathbf{I}_N$. The big-O notation is given by $O(\cdot)$. A diagonal matrix with diagonal entries $x_1,\dots,x_N$ is denoted as $\mathrm{diag}(x_1,\dots,x_N)$.


\section{Fundamental Physics-based Hardware and Signal Models} \label{sec:model}

In this section, we introduce the fundamental physics-based hardware and signal models for pinching antennas. Specifically, we model a pinching antenna as an open-ended directional coupler, which facilitates the adjustment of radiation characteristics and simplifies signal modeling. We then employ coupled-mode theory to characterize the relationship between the signal within the waveguide and the signal radiated by the pinching antenna.

\begin{figure}[t!]
    \centering
    \includegraphics[width=0.45\textwidth]{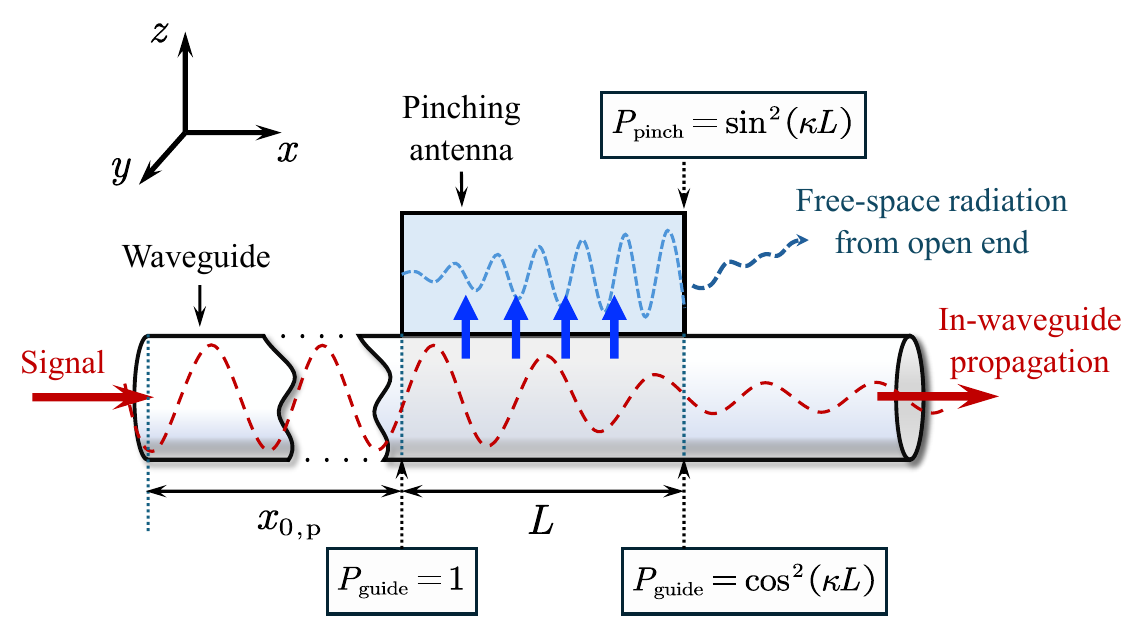}
    \caption{Schematic illustration of pinching antennas operating as an open-ended directional waveguide coupler.}
    \label{physical_model}
  \end{figure} 

\subsection{Physics-based Hardware Model}

The core principle of pinching antennas relies on the phenomenon where a portion of the EM waves propagating through a dielectric waveguide is induced into an adjacent dielectric material (i.e., a pinching antenna) when the two are placed in close proximity\footnote{\textcolor{black}{It is possible to apply pinching antennas on both sides of a specific location along the waveguide, which, however, would significantly increase the complexity of the model. In this paper, we restrict our analysis to a single pinching antenna per location to maintain a clear and tractable framework.}}. To accurately model the signal radiated by the pinching antenna, it is essential to characterize the EM fields within both the waveguide and the pinching antenna, as well as to understand the interaction between these fields. 

Consider a dielectric waveguide with an effective refractive index $n_{\mathrm{g}}$. A signal with a free-space wavelength $\lambda$ is introduced into the waveguide and propagates along the $x$-axis, as shown in Fig. \ref{physical_model}. The electric field distribution of the EM wave within the waveguide can be expressed as:
\begin{equation} \label{EM_model_waveguide}
    \mathbf{E}_{\mathrm{guide}}(x,y,z) = \mathbf{D}_{\mathrm{guide}}(y,z) e^{-j \beta_{\mathrm{g}} x} s_{\mathrm{p}},
\end{equation}
where $\mathbf{D}_{\mathrm{guide}}(y,z) \in \mathbb{C}^{3 \times 1}$ represents the transverse field distribution of the guided mode, $\beta_{\mathrm{g}} = \frac{2 \pi n_{\mathrm{g}}}{\lambda}$ is the propagation constant of the waveguide, and $s_{\mathrm{p}} \in \mathbb{C}$ denotes the phase-shifted communication signal modulated onto the EM wave. Let $x_{0, \mathrm{p}}$ denote the distance that the signal has propagated within the waveguide prior to the starting point of coupling, as shown in Fig. \ref{physical_model}. The signal $s_{\mathrm{p}}$ can be expressed as:
\begin{equation}
    s_{\mathrm{p}} = e^{-j \beta_{\mathrm{g}} x_{0, \mathrm{p}}} c_0,
\end{equation}
where $c_0$ represents the original communication signal. Note that the in-waveguide propagation loss is omitted in the above formulas due to its negligible impact on the system performance \cite{ding2024flexible}. 

We model the pinching antenna as an open-ended directional waveguide coupler, where EM waves can be radiated from one end of the pinching antenna with minimal reflection by optimizing waveguide's aperture size, shape, and termination impedance \cite{gardiol1985open}. For simplicity, we assume ideal full radiation from the open end of the pinching antenna with no reflection. When the pinching antenna is placed in proximity to (or “pinched” against) the main waveguide, coupling occurs, generating an EM field within the pinching antenna. Let $n_{\mathrm{p}}$ denote the effective refractive index of the pinching antenna. The electric field within the pinching antenna is
\begin{equation}
    \mathbf{E}_{\mathrm{pinch}}(x,y,z) = \mathbf{D}_{\mathrm{pinch}}(y,z) e^{-j \beta_{\mathrm{p}} x} s_{\mathrm{p}},
\end{equation}
where $\mathbf{D}_{\mathrm{pinch}}(y,z) \in \mathbb{C}^{3 \times 1}$ represents the transverse field distribution, and $\beta_{\mathrm{p}} = \frac{2 \pi n_{\mathrm{p}}}{\lambda}$ is the propagation constant of the pinching antenna.

Based on coupled-mode theory, the total EM field within the waveguide and the pinching antenna can be expressed as a weighted sum of their respective individual fields. Let $\mathbf{E}$ denote the overall electric field, which can be written as:
\begin{equation}
    \mathbf{E} = A(x) \mathbf{E}_{\mathrm{guide}} + B(x) \mathbf{E}_{\mathrm{pinch}}.
\end{equation}
By substituting the expressions for the electric field $\mathbf{E}$ and the corresponding magnetic field into Maxwell's equations, the following coupled differential equations for $A(x)$ and $B(x)$ are obtained \cite{okamoto2010fundamentals}:
\begin{align} \label{diff_condition_A}
    \frac{d A(x)}{d x} &= -j \kappa B(x) e^{-j \Delta \beta x}, \\
    \label{diff_condition_B}
    \frac{d B(x)}{d x} &= -j \kappa A(x) e^{j \Delta \beta x},
\end{align}
where $\kappa \in \mathbb{R}$ is the mode coupling coefficient, determined by the transverse field distributions $\mathbf{D}_{\mathrm{guide}}(y,z)$ and $\mathbf{D}_{\mathrm{pinch}}(y,z)$, and $\Delta \beta = \beta_{\mathrm{p}} - \beta_{\mathrm{g}}$ represents the difference between the propagation constants of the waveguide and the pinching antenna. Since the signal is initially introduced into the waveguide, the following initial conditions hold:
\begin{align}
    A(0) = 1, \quad B(0) = 0.
\end{align}
Solving the differential equations \eqref{diff_condition_A} and \eqref{diff_condition_B} under these initial conditions yields the following expressions for $A(x)$ and $B(x)$:
\begin{align}
    A(x) &= \left( \cos \left(\phi x \right) + \frac{j \Delta \beta}{2 \phi} \sin(\phi x) \right) e^{-j \Delta \beta x/2}, \\
    B(x) &= - \frac{j \kappa}{\phi} \sin(\phi x) e^{j \Delta \beta x/2},
\end{align}
where $\phi = \sqrt{\kappa^2 + \Delta \beta^2/4}$. Consequently, the total power of the EM field within the waveguide and the pinching antenna is given by:
\begin{align}
    P_{\mathrm{guide}}(x) &= \left| A(x) \right|^2 = 1 - \left( \frac{\kappa}{\phi} \right)^2 \sin^2(\phi x), \\
    P_{\mathrm{pinch}}(x) &= \left| B(x) \right|^2 = \left( \frac{\kappa}{\phi} \right)^2 \sin^2(\phi x).
\end{align}
From these expressions, it is evident that a maximum fraction of $\left( \kappa/\phi \right)^2$ of the total power can be transferred to the pinching antenna. \textcolor{black}{In the rest of this paper, we consider a special case where the waveguide and the pinching antenna have the same effective refractive index (i.e., $\beta_{\mathrm{g}} = \beta_{\mathrm{p}}$), where we have $\Delta \beta = 0$ and $\phi = \kappa$.} Under these conditions, the expressions for $A(x)$ and $B(x)$ can be simplified to:
\begin{align} \label{simplified_power_exchange}
    A(x) = \cos(\kappa x), \quad B(x) = -j \sin(\kappa x).
\end{align}
Utilizing the simplified power exchange coefficients in \eqref{simplified_power_exchange} for the case $\beta_{\mathrm{g}} = \beta_{\mathrm{p}}$, the signal radiated from the open end of the pinching antenna can be expressed as
\begin{align} \label{EM_model_rad}
    \mathbf{E}_{\mathrm{rad}}(y,z) &=  B(L) \mathbf{E}_{\mathrm{pinch}}(L,y,z) \nonumber \\
    &= -j \mathbf{D}_{\mathrm{pinch}}(y,z) \sin(\kappa L) e^{-j \beta_{\mathrm{g}} x_{\mathrm{p}}} c_0,
\end{align}
where $x_{\mathrm{p}}$ represents the effective position of the pinching antenna and is defined as:
\begin{equation}
    x_{\mathrm{p}} = x_{0, \mathrm{p}} + L.
\end{equation}
Once the field in the pinching antenna radiates into free space, the power exchange between the waveguide and the pinching antenna ceases at $x = L$. The remaining EM wave propagating within the waveguide is given by:
\begin{align} \label{EM_model_guide_remain}
    \tilde{\mathbf{E}}_{\mathrm{guide}}(x,y,z) &= A(L) \mathbf{E}_{\mathrm{guide}}(L,y,z) \nonumber \\
    &= \mathbf{D}_{\mathrm{guide}}(y,z) \cos(\kappa L) e^{-j \beta_{\mathrm{g}} (x + x_{\mathrm{p}})} c_0.
\end{align}

\begin{remark}
    \normalfont
    \emph{(Power Relationship)} Based on the above physics model, the power of the signals within the waveguide and radiated by the pinching antenna is determined by the coupling length $L$. In the case of $\beta_{\mathrm{g}} = \beta_{\mathrm{p}}$, a complete power transfer to the pinching antenna becomes achievable, enabling full signal radiation, which is achieved when $L = \pi/(2\kappa)$ according to \eqref{simplified_power_exchange}. However, this strategy is applicable for the case with a single pinching antennas, but not for the case with multiple pinching antennas on the same waveguide. For the latter case, the radiation power from each pinching antenna can be adjusted by modifying the coupling length $L$. In other words, the transmit powers of the multiple pinching antennas can be controlled, which leads to extra degrees of freedom (DoFs) for the system design; howerver, this would likely lead to active pinching antenna design and additional system complexity. Therefore, this paper focuses on a purely passive pinching antenna design with a preconfigured coupling length.
\end{remark}

\subsection{Signal Model}
\textcolor{black}{In \eqref{EM_model_rad} and \eqref{EM_model_guide_remain}, we derived the physical models describing the EM wave behavior in the PASS system, assuming identical effective refractive indices for the waveguides and pinching antennas. Building upon these foundational physics models}, we can now formulate simplified signal models commonly employed in the wireless communication system design. In the following, we begin by considering the scenario with a single pinching antenna coupled to the waveguide and subsequently extend the model to cases involving multiple pinching antennas.

\subsubsection{A Single Pinching Antenna}
Assume that a waveguide is deployed along the $x$-axis, with its $y$- and $z$-coordinates denoted by $y_{\mathrm{g}}$ and $z_{\mathrm{g}}$, respectively. A user is located on the ground at the position $\mathbf{r} = [x_{\mathrm{u}}, y_{\mathrm{u}}, 0]^T$. To serve this user, a pinching antenna is attached to the waveguide at the position $\mathbf{p} = [x_{\mathrm{p}}, y_{\mathrm{g}}, z_{\mathrm{g}}]^T$. Let $c_0$ represent the signal fed into the waveguide. According to \eqref{EM_model_rad}, the signal radiated from the pinching antenna is given by
\begin{equation}
s_{\mathrm{rad}} = \sin(\kappa L) e^{-j \beta_{\mathrm{g}} x_{\mathrm{p}}} c_0,
\end{equation}
The radiated signal propagates through free space to reach the user, where attenuation due to free-space path loss must be considered, leading to the following signal model\footnote{\textcolor{black}{In our signal model, we consider only LoS path in the free space due to two reasons. First, it is important to note that dielectric waveguides typically operate in high-frequency bands, such as the millimeter-wave bands \cite{suzuki2022pinching}. At these high frequencies, the LoS path can be 20 dB stronger than NLoS paths \cite{6363891}, making the impact of NLoS components almost negligible. Second, even if NLoS paths are relevant, their path loss exponents are generally higher than that of the LoS path. Thus, the performance benefits provided by PASS can be even more significant, since it effectively reduce the transmission distance.}}:
\begin{align}
y = & \frac{\eta e^{-j \beta_0 r}}{r} s_{\mathrm{rad}} + n \nonumber \\
= & \underbrace{\frac{\eta e^{-j \beta_0 r}}{r}}_{\scriptstyle \text{high-loss free-space} \atop \scriptstyle \text{propagation}} \!\!\! \times \underbrace{ \vphantom{\frac{\eta e^{-j \beta_0 r}}{r}} \sin(\kappa L) e^{-j \beta_{\mathrm{g}} x_{\mathrm{p}}}}_{\scriptstyle \text{nearly lossless in-waveguide} \atop \scriptstyle  \text{propagation}}  c_0 + n,
\end{align}
where $\eta \in \mathbb{R}$ is the channel gain accounting for the free-space path loss factor and the radiation pattern of the pinching antennas, $\beta_0 = \frac{2 \pi}{\lambda}$ is the propagation constant in free space, $r$ is the distance between the pinching antenna and the user, given by
\begin{equation}
r = \| \mathbf{r} - \mathbf{p} \| = \sqrt{(x_{\mathrm{p}} - x_{\mathrm{u}})^2 + \omega},
\end{equation}
with $\omega = (y_{\mathrm{g}} - y_{\mathrm{u}})^2 + z_{\mathrm{g}}^2$, and $n \sim \mathcal{CN}(0 ,\sigma^2)$ denotes the additive white Gaussian noise.

\subsubsection{Multiple Pinching Antennas}

Now we consider a scenario where $M$ pinching antennas are pinched sequentially on the waveguide. Let $\mathbf{p}_m = [x_{\mathrm{p},m}, y_{\mathrm{g}}, z_{\mathrm{g}}]^T$ and $L_m$ denote the position and the length of the $m$-th pinching antenna, respectively. According to \eqref{EM_model_rad} and \eqref{EM_model_guide_remain}, the power radiated by the $m$-th pinching antenna is influenced by the power exchange coefficients of all preceding pinching antennas. Consequently, the signal radiated from the $m$-th pinching antenna can be expressed as: 
\begin{align}
    s_{\mathrm{rad}, m} =&\sin(\kappa L_m) \prod_{i=1}^{m-1} \cos(\kappa L_i) e^{-j \beta_{\mathrm{g}} x_{\mathrm{p},m}} c_0 \nonumber \\
    = & \delta_m \prod_{i=1}^{m-1} \sqrt{1 - \delta_i^2} e^{-j \beta_{\mathrm{g}} x_{\mathrm{p},m}} c_0,
\end{align} 
where we define $\delta_m \triangleq \sin(\kappa L_m)$. 
Here, we consider two simplified but useful signal radiation model for multiple pinching antennas:
\begin{itemize}
    \item \textbf{Equal Power Model:} \textcolor{black}{In this model, we assume that the length $L_m$ of each pinching antenna is adjusted so that each antenna radiates an equal proportion of the total power. Specifically, this is expressed as
    \begin{align}
        &\delta_m \prod_{i=1}^{m-1} \sqrt{1 - \delta_i^2} = \sqrt{\delta_{\mathrm{eq}}}, \quad \forall m, \nonumber \\
        \Leftrightarrow \quad & \delta_m = \sqrt{\frac{\delta_{\mathrm{eq}}}{1 - (m-1)\delta_{\mathrm{eq}}}}, \quad \forall m,
    \end{align}
    where $0 < \delta_{\mathrm{eq}} \le \frac{1}{M}$ is the equal-power ratio.} Under this condition, the radiated signal from the $m$-th pinching antenna is simplified into
    \begin{equation}
        s_{\mathrm{rad},m} = \sqrt{\delta_{\mathrm{eq}}}  e^{-j \beta_{\mathrm{g}} x_{\mathrm{p},m}} c_0.
    \end{equation}  
    This model ensures that each pinching antenna radiates with the same efficiency, making it useful for obtaining insight to the performance of PASS as discussed in \cite{ding2024flexible}. However, achieving this equal-power distribution requires that each pinching antenna be manufactured with a different length, which increases the hardware cost.
    \item \textbf{Proportional Power Model:} In this model, we assume that the length of each pinching antenna is manufactured to be the same, i.e., $L_m = L, \forall m$. Consequently, each pinching antenna radiates the same ratio of the remaining power within the waveguide. Under this condition, the radiated signal from the $m$-th pinching antenna becomes
    \begin{equation}
        s_{\mathrm{rad},m} = \delta \left(\sqrt{1 - \delta^2}\right)^{m-1} e^{-j \beta_{\mathrm{g}} x_{\mathrm{p},m}} c_0,
    \end{equation}
    where $\delta = \sin(\kappa L)$. Compared to the equal-power model, this equal-ratio model significantly reduces hardware costs, as all pinching antennas can be uniformly manufactured with the same length.
\end{itemize}

Based on the above modeling, the signal received from all pinching antennas at the user can be expressed as 
\begin{align} \label{multiple_basis_model}
    y = &\sum_{m=1}^M \frac{\eta e^{-j \beta_0 r_m}}{r_m} s_{\mathrm{rad}, m} + n = \mathbf{h}^H(\mathbf{x}) \mathbf{g}(\mathbf{x}) c_0 + n.
\end{align}
Here, $r_m = \|\mathbf{r} - \mathbf{p}_m\|$ is the distance between the $m$-th pinching antenna and the user. $\mathbf{h}(\mathbf{x}) \in \mathbb{C}^{M \times 1}$ represents the free-space channel vector between all pinching antennas and the user, given by 
\begin{equation} \label{basic_channel_model}
    \mathbf{h}(\mathbf{x}) = \left[\frac{\eta e^{-j \beta_0 r_1}}{r_1},\dots,\frac{\eta e^{-j \beta_0 r_M}}{r_M}  \right]^H,
\end{equation}
which is a function of the pinching antenna positions $\mathbf{x} = [x_{\mathrm{p},1},\dots,x_{\mathrm{p},M}]^T$. $\mathbf{g}(\mathbf{x}) \in \mathbb{C}^{M \times 1}$ denotes the in-waveguide channel vector, given by 
\begin{equation} \label{basic_pinching_beamforming_model}
    \mathbf{g}(\mathbf{x}) = \left[ \alpha_1 e^{-j \beta_{\mathrm{g}} x_{\mathrm{p},1}},\dots,\alpha_M e^{-j \beta_{\mathrm{g}} x_{\mathrm{p},M}} \right]^T,
\end{equation} 
where $\alpha_m = \sqrt{\delta_{\mathrm{eq}}}$ for the equal power model, while $\alpha_m = \delta(\sqrt{1 - \delta^2})^{m-1}$ for the proportional power model.  

\begin{remark}
    \normalfont
    \emph{(Pinching Beamforming)} \textcolor{black}{Pinching antennas extract signals from waveguides through EM coupling with no physical connection to the waveguide. Therefore, it is possible to dynamically reposition the pinching antennas. For instance, a pinching antenna can be made movable by placing it on a pre-installed track parallel to the waveguide, which is similar to how a motorized camera slider operates \cite{ding2024flexible}.} It can be observed from \eqref{multiple_basis_model}-\eqref{basic_pinching_beamforming_model} that adjusting the positions of the pinching antennas, determined by $\mathbf{x}$, alters the phase and the large-scale path loss (amplitude) of the signal received by the user. Analogous to conventional multi-antenna beamforming, the signals from multiple pinching antennas can be combined either constructively or destructively at the user by carefully optimizing the positions of the pinching antennas. This new capability for signal reconfiguration introduced by PASS is referred to as \emph{pinching beamforming}.
    
\end{remark}


\begin{figure}[t!]
    \centering
    \includegraphics[width=0.45\textwidth]{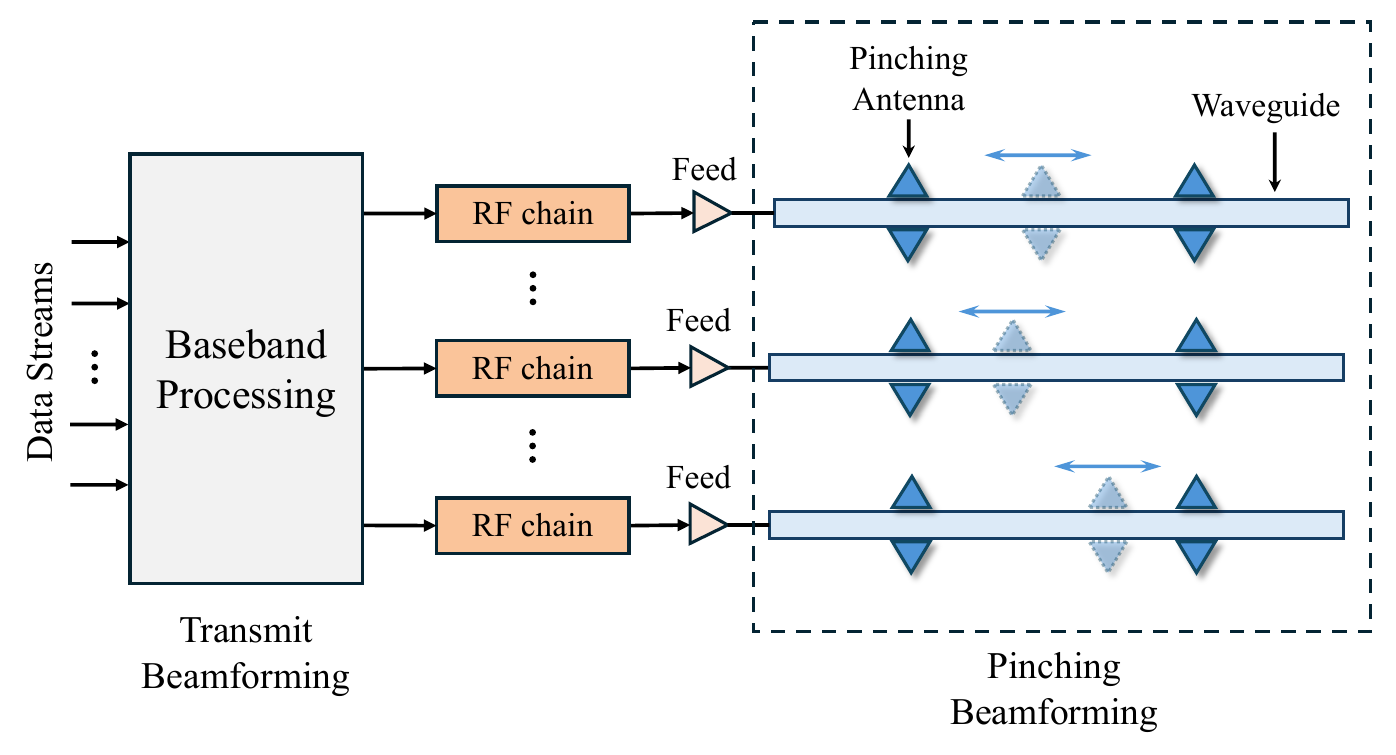}
    \caption{Joint transmit and pinching beamforming architecture.}
    \label{beamforming_structure}
\end{figure}

\section{Joint Transmit and Pinching Beamforming} \label{sec:beamforing}

Based on the signal model, this section studies a PASS enabled downlink multi-user communication system and the resultant joint transmit and pinching beamforming optimization problem.

\subsection{System Model}

We consider a downlink PASS consisting of $N$ dielectric waveguides serving $K$ communication users, where $N \ge K$. The joint transmit and pinching beamforming architecture is depicted in Fig. \ref{beamforming_structure}, where each waveguide is fed by a dedicated RF chain and incorporates $M$ pinching antennas. Let $\mathbf{r}_k = [x_{\mathrm{u},k}, y_{\mathrm{u},k}, z_{\mathrm{u},k}]^T \in \mathbb{R}^{3 \times 1}$ denote the user $k$'s position and $\mathbf{p}_{n,m} = [x_{n,m}, y_{\mathrm{g},n}, z_{\mathrm{g},n}]^T \in \mathbb{R}^{3 \times 1}$ denote the position of the $m$-th pinching antenna on the $n$-th waveguide.

\subsubsection{Transmit Signal} 
Let $\mathbf{c} = [c_1,\dots,c_K]^T \in \mathbb{C}^{K \times 1}$ denote the complex information symbols with $c_k$ intended for user $k$, which satisfies $\mathbb{E} \{\mathbf{c} \mathbf{c}^H\} = \mathbf{I}_K$. The signal after the transmit beamforming is given by
\begin{equation}
    \mathbf{s} = \mathbf{W}  \mathbf{c} = \sum_{k=1}^K \mathbf{w}_k c_k,
\end{equation}
where $\mathbf{W} = [\mathbf{w}_1,\dots,\mathbf{w}_K] \in \mathbb{C}^{N \times K}$ denote the overall transmit beamforming matrix and $\mathbf{w}_k \in \mathbb{C}^{N \times 1}$ is the beamforming vector for user $k$. 

\subsubsection{Receive Signal}
Let $s_n$ denote the $n$-th entry of the signal $\mathbf{s}$, which represents the signal introduced into the $n$-th waveguide. According to \eqref{multiple_basis_model}, the signal received at user $k$ can be expressed as 
\begin{equation} \label{received_signal_1}
    y_k = \sum_{n=1}^N \mathbf{h}_k^H(\mathbf{x}_n) \mathbf{g}(\mathbf{x}_n) s_n + n_k,
\end{equation} 
where $\mathbf{h}_k(\mathbf{x}_n) \in \mathbb{C}^{M \times 1}$ and $\mathbf{g}(\mathbf{x}_n) \in \mathbb{C}^{M \times 1}$ are the free-space and in-waveguide channel vectors, respectively, and $n_k \sim \mathcal{CN}(0, \sigma_k^2)$ denotes the additive white Gaussian noise. According to \eqref{basic_channel_model} and \eqref{basic_pinching_beamforming_model}, we have
\begin{align}
    \mathbf{h}_k(\mathbf{x}_n) = &\left[\frac{\eta e^{-j \beta_0 r_{k,n,1}}}{r_{k,n,1}},\dots,\frac{\eta e^{-j \beta_0 r_{k,n,M}}}{r_{k,n,M}}  \right]^H, \\
    \mathbf{g}(\mathbf{x}_n) = &\left[ \alpha_1 e^{-j \beta_{\mathrm{g}} x_{n, 1}},\dots,\alpha_M e^{-j \beta_{\mathrm{g}} x_{n, M}} \right]^T,
\end{align}
where $\mathbf{x}_n = [x_{n,1},\dots,x_{n,M}]^T$ denotes the $x$-axis position vector for all pinching antennas on the $n$-th waveguide, and $r_{k,n,m}$ denotes the distance between user $k$ and the $m$-th pinching antenna on the $n$-th waveguide, given by     
\begin{align}
    r_{k,n,m} = &\| \mathbf{r}_k - \mathbf{p}_{n,m} \| = \sqrt{ (x_{n,m} - x_{\mathrm{u},k})^2 + \omega_{k,n} },
\end{align}
with $\omega_{k,n} = (y_{\mathrm{g},n} - y_{\mathrm{u},k})^2 + z_{\mathrm{g},n}^2$. The expression \eqref{received_signal_1} for the signal received at user $k$ can be rewritten more compactly as follows:
\begin{align}
    y_k = & \mathbf{h}_k^H (\mathbf{X}) \mathbf{G}(\mathbf{X}) \mathbf{s} + n_k \nonumber \\
    = & \underbrace{\vphantom{\sum_{i=1, i\neq k}^K} \mathbf{h}_k^H (\mathbf{X}) \mathbf{G}(\mathbf{X}) \mathbf{w}_k c_k}_{\text{desired signal}} +  \underbrace{\sum_{i=1, i\neq k}^K \mathbf{h}_k^H (\mathbf{X}) \mathbf{G}(\mathbf{X}) \mathbf{w}_i c_i}_{\text{inter-user interference}} + n_k,
\end{align}
where $\mathbf{X} = [\mathbf{x}_1,\dots,\mathbf{x}_N] \in \mathbb{R}^{M \times N}$ consists of the positions of all pinching antennas, $\mathbf{h}_k (\mathbf{X}) = \big[\mathbf{h}_k^T(\mathbf{x}_1),\dots,\mathbf{h}_k^T(\mathbf{x}_N)\big]^T \in \mathbb{C}^{NM \times 1}$ denotes the overall free-space channel vector for user $k$, and $\mathbf{G}(\mathbf{X}) \in \mathbb{C}^{NM \times N}$ denotes the overall in-waveguide channel matrix, given by 
\begin{equation}
    \mathbf{G}(\mathbf{X}) = \begin{bmatrix}
        \mathbf{g}(\mathbf{x}_1) & \mathbf{0} & \cdots & \mathbf{0}\\
        \mathbf{0} & \mathbf{g}(\mathbf{x}_2) & \cdots & \mathbf{0} \\
        \vdots & \vdots & \ddots &  \vdots \\
        \mathbf{0} & \mathbf{0} & \cdots & \mathbf{g}(\mathbf{x}_N)
    \end{bmatrix}.
\end{equation}
Accordingly, the signal-to-interference-plus-noise ratio (SINR) for user $k$ to decode its own signal is given by
\begin{equation} \label{SINR_formula}
    \mathrm{SINR}_k(\mathbf{W}, \mathbf{X}) = \frac{ \left| \mathbf{h}_k^H(\mathbf{X}) \mathbf{G}(\mathbf{X}) \mathbf{w}_k \right|^2 }{\sum_{i = 1, i \neq k}^K \left| \mathbf{h}_k^H(\mathbf{X}) \mathbf{G}(\mathbf{X}) \mathbf{w}_i \right|^2 + \sigma_k^2}.
\end{equation}

\subsection{Problem Formulation}

In this paper, we aim to minimize the total transmission power at the BS by jointly optimizing the transmit beamforming at the BS and the pinching beamforming facilitated by the pinching antennas, under the constraint of meeting the individual minimum SINR requirements for all users. Let $\mathbf{W} = [\mathbf{w}_1,\dots,\mathbf{w}_K]$ and $\mathbf{H}(\mathbf{X}) = [\mathbf{h}_1(\mathbf{X}),\dots,\mathbf{h}_K(\mathbf{X})]$. The problem can be formulated as
\begin{subequations} \label{hybrid_beamforming_problem}
    \begin{align}
        \min_{\mathbf{W}, \mathbf{X}} \quad & \sum_{k=1}^K \|\mathbf{w}_k\|^2 \\
        \mathrm{s.t.} \quad & \mathrm{SINR}_k(\mathbf{W}, \mathbf{X}) \ge \gamma_k, \forall k, \\
        \label{theta_constraint_1}
        & x_{n,m} - x_{n,m-1} \ge \Delta x, \forall n, m\neq 1, \\
        \label{theta_constraint_2}
        & x_{n,m} \in \mathcal{S},
    \end{align}
\end{subequations}
where $\gamma_k > 0$ represents the minimum SINR requirement of user $k$, $\Delta x$ is the minimum spacing required to prevent mutual coupling between the pinching antennas, and $\mathcal{S}$ denotes the feasible set of the positions of the pinching antennas. In this paper, we consider both the continuous and discrete activation scenarios for the pinching antennas, resulting in the following two distinct types of feasible sets:
\begin{itemize}
    \item \textbf{Continuous Activation:} In this case, the pinching antennas can be activated at arbitrary position on the waveguide, leading to the following the feasible set $\mathcal{S}$:
    \begin{equation}
        \mathcal{S} = [0, x_{\max}],
    \end{equation} 
    where $x_{\max} > 0$ represents the maximum deployment range of the pinching antennas. This is the ideal case for pinching antenna deployment, which yields the performance upper bound.
    \item \textbf{Discrete Activation:} In this case, the pinching antennas can be activated at preconfigured discrete positions only, leading to the following feasible set $\mathcal{S}$: 
    \begin{equation}
        \mathcal{S} = \left\{0, \frac{x_{\max}}{Q-1}, \dots,x_{\max} \right\},
    \end{equation} 
    where $Q$ denotes the number of discrete positions available for deploying the pinching antennas. Compared to the continuous activation, the discrete activation can reduce the hardware complexity and is thus more practical. 
\end{itemize}

\begin{figure}[t!]
    \centering
    \includegraphics[width=0.45\textwidth]{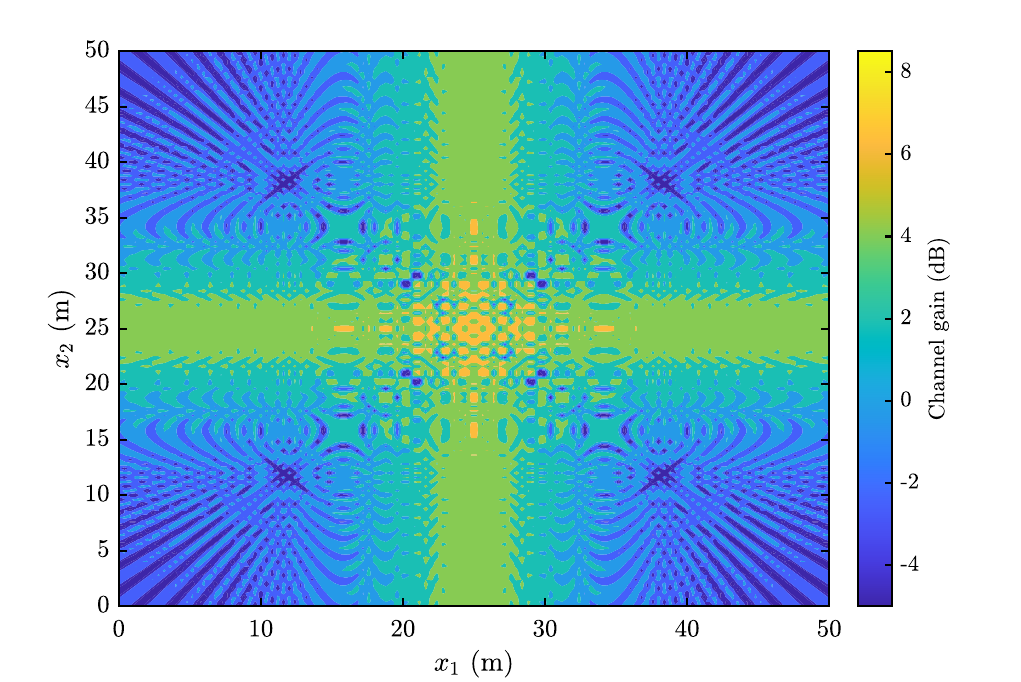}
    \caption{Illustration of the highly multimodal channel gain with respect to the pinching antenna positions in a “simple” scenario where $N = 1$, $M=2$, and $K=1$. More specifically, $x_1$ and $x_2$ are the positions of the two pinching antennas on the waveguide.}
    \label{fig_multimodal}
\end{figure}

Although the objective function of problem \eqref{hybrid_beamforming_problem} is convex, solving it remains challenging due to the coupling between the transmit beamforming matrix $\mathbf{W}$ and the antenna position matrix $\mathbf{X}$, as well as the highly coupled and nonlinear dependency of $\mathbf{H}(\mathbf{X})$ and $\mathbf{G}(\mathbf{X})$ on the position matrix $\mathbf{X}$ of the pinching antennas. It is worth noting that a similar antenna position optimization problem arises in the beamforming design for fluid and movable antenna systems \cite{zhu2023movable}, where a common approach to solving this problem is gradient descent. However, this method is not suitable for optimizing the positions of pinching antennas for two key reasons. First, the optimization problem in \eqref{hybrid_beamforming_problem} is highly multimodal, meaning it contains a large number of local optima, as illustrated in Fig. \ref{fig_multimodal}. Second, the performance gap between local optima can be extremely large due to the significant oscillations in free-space path loss, which result from the wide deployment range of pinching antennas. To overcome these challenges, two optimization algorithms are proposed in the following: 1) the penalty-based alternating optimization algorithm and 2) the ZF-based low-complexity algorithm.

\subsection{Penalty-based Alternating Optimization Algorithm}

In this section, we propose a penalty-based alternating optimization algorithm to address the joint beamforming optimization problem \eqref{hybrid_beamforming_problem}. The key idea of alternating optimization is to alternatively fix one subset of variables, either the transmit beamforming matrix $\mathbf{W}$ or the antenna position matrix $\mathbf{X}$, and optimize the objective function with respect to the other. 

However, it is important to note that the objective function in \eqref{hybrid_beamforming_problem} depends only on the matrix $\mathbf{W}$. As a result, when optimizing $\mathbf{X}$ while keeping $\mathbf{W}$ fixed, the corresponding subproblem with respect to $\mathbf{X}$ becomes a feasibility-checking problem, i.e., with the objective function being a constant. Consequently, directly solving \eqref{hybrid_beamforming_problem} using alternating optimization does not ensure sequential minimization of the objective function across iterations, which can lead to poor convergence performance. To get rid of this issue, we define $\mathbf{v}_k = \mathbf{w}_k/\sqrt{P}$, where $P = \sum_{k=1}^K \|\mathbf{w}_k\|^2$, and reformulate problem \eqref{hybrid_beamforming_problem} into the following equivalent form:   
\begin{subequations} \label{joint_beamforming_problem_2}
    \begin{align}
        \min_{\mathbf{V}, \mathbf{X}, P > 0} \quad & P \\
        \mathrm{s.t.} \quad & \overline{\mathrm{SINR}}_k(\mathbf{V}, \mathbf{X}, P)  \ge \gamma_k, \forall k, \\
        \label{unit_power_constraint}
        & \sum_{k=1}^K \| \mathbf{v}_k \|^2 = 1, \\
        & \eqref{theta_constraint_1}, \eqref{theta_constraint_2},
    \end{align}
\end{subequations}
where 
\begin{equation}
    \overline{\mathrm{SINR}}_k(\mathbf{V}, \mathbf{X}, P) = \frac{ P \left| \mathbf{h}_k^H(\mathbf{X}) \mathbf{G}(\mathbf{X}) \mathbf{v}_k \right|^2 }{\sum_{i = 1, i \neq k}^K P \left| \mathbf{h}_k^H(\mathbf{X}) \mathbf{G}(\mathbf{X}) \mathbf{v}_i \right|^2 + \sigma_k^2 }.
\end{equation}
As can be observed, in problem \eqref{joint_beamforming_problem_2}, the objective function is related to both $\mathbf{W}$ and $\mathbf{X}$ through the SINR constraint, which facilitates the convergence when the alternating optimization technique is applied. 

The primary challenge in solving the reformulated problem \eqref{joint_beamforming_problem_2} arises from the highly coupled and nonlinear nature of $\mathbf{H}(\mathbf{X})$ and $\mathbf{G}(\mathbf{X})$ within the SINR constraints. The penalty method is a well-established approach in the literature for addressing such coupling issues by introducing appropriate equality constraints and incorporating them into the objective function \cite{liu2018mu, shi2020penalty, wu2020joint}. To effectively apply the penalty method, we first define the following decomposition of the matrices $\mathbf{H}(\mathbf{X})$ and $\mathbf{G}(\mathbf{X})$:
\begin{align} \label{PH_formula}
    \mathbf{G}^H(\mathbf{X}) \mathbf{H}(\mathbf{X})  
    & = \begin{bmatrix}
        \mathbf{g}^H(\mathbf{x}_1) \mathbf{h}_1(\mathbf{x}_1) & \!\! \cdots \!\! & \mathbf{g}^H(\mathbf{x}_1) \mathbf{h}_K(\mathbf{x}_1) \\ \vdots & \!\! \ddots \!\!  & \vdots \\
        \mathbf{g}^H(\mathbf{x}_N) \mathbf{h}_1(\mathbf{x}_N) & \!\! \cdots \!\! & \mathbf{g}^H(\mathbf{x}_N) \mathbf{h}_K(\mathbf{x}_N)
    \end{bmatrix} \nonumber \\ &= \sum_{m=1}^M \mathbf{\Phi}_m(\mathbf{X}),
\end{align}
where the entry of $\mathbf{\Phi}_m(\mathbf{X})$ in the $n$-th row and $k$-th column is given by 
\begin{equation}
   \left[ \mathbf{\Phi}_m(\mathbf{X}) \right]_{n,k} = \frac{\eta \alpha_m }{r_{k,n,m}} e^{j\left( \beta_0 r_{k,n,m} + \beta_{\mathrm{g}} x_{n,m} \right)}.
\end{equation}
It will be shown in the following that the above decomposition significantly reduces the complexity of optimizing the pinching antenna positions and addresses the issue of multimodality. Leveraging this decomposition, problem \eqref{joint_beamforming_problem_2} can be reformulated into the following equivalent form:
\begin{subequations} \label{joint_beamforming_problem_3}
    \begin{align}
        \min_{\mathbf{U}, \mathbf{U}_m, \mathbf{V}, \mathbf{X}, P > 0} \quad & P \\
        \label{joint_beamforming_problem_3_0}
        \mathrm{s.t.} \quad & \mathbf{U} = \sum_{m=1}^M \mathbf{U}_m, \mathbf{U}_m = \mathbf{\Phi}_m(\mathbf{X}), \forall m, \\
        \label{joint_beamforming_problem_3_1}
        & \overline{\overline{\mathrm{SINR}}}_k(\mathbf{V}, \mathbf{U}, P)  \ge \gamma_k, \forall k, \\
        & \eqref{theta_constraint_1}, \eqref{theta_constraint_2}, \eqref{unit_power_constraint},
    \end{align}
\end{subequations}
where 
\begin{equation}
    \overline{\overline{\mathrm{SINR}}}_k(\mathbf{V}, \mathbf{X}, P) = \frac{ P \left| \mathbf{u}_k^H \mathbf{v}_k \right|^2 }{\sum_{i = 1, i \neq k}^K P \left| \mathbf{u}_k^H \mathbf{v}_i \right|^2 + \sigma_k^2 }.
\end{equation}
In this new problem, $\mathbf{U} = [\mathbf{u}_1,\dots,\mathbf{u}_k] \in \mathbb{C}^{NM \times K}$ is an auxiliary channel matrix. By integrating the equality constrains into the objective function as penalty terms, problem \eqref{joint_beamforming_problem_3} becomes
\begin{subequations} \label{joint_beamforming_problem_4}
    \begin{align}
        \min_{\mathbf{U}, \mathbf{U}_m, \mathbf{V}, \mathbf{X}, P > 0} \quad P &+ \frac{1}{\rho} \| \mathbf{U} - \sum_{m=1}^M \mathbf{U}_m \|_F^2 \nonumber \\
        & + \frac{1}{\rho} \sum_{m=1}^M \| \mathbf{U}_m - \mathbf{\Phi}_m(\mathbf{X}) \|_F^2 \\
        \mathrm{s.t.} \quad &\eqref{theta_constraint_1}, \eqref{theta_constraint_2}, \eqref{unit_power_constraint}, \eqref{joint_beamforming_problem_3_1}, 
    \end{align}
\end{subequations}
where $\rho > 0$ is the penalty factor. In particular, when $\rho \to 0$, the penalty terms can be exactly zero, ensuring that the original equality constraints are strictly satisfied. However, this approach renders the transmit power $P$ a dummy objective function, leading to significant performance degradation. To mitigate this issue, the penalty factor is initially set to a large value to ensure sufficient minimization of the transmit power. It is then gradually reduced, allowing the penalty terms to converge to zero, i.e., forcing the auxiliary channel matrix $\mathbf{U}$ to follow the channel structure in PASS. Therefore, the penalty method typically has a double-loop optimization structure. In the outer loop, the penalty factor is progressively decreased, while in the inner loop, the optimization variables are updated using an alternating optimization approach. The solutions for each subset of optimization variables, while keeping the other subsets fixed, are presented in the following.

\subsubsection{Subproblem With Respect to $\mathbf{V}$} 
Based on the definition $\mathbf{v}_k = \mathbf{w}_k/\sqrt{P}$ and $P = \sum_{k=1}^K \|\mathbf{w}_k\|^2$, this subproblem can be rewritten as
\begin{subequations} \label{subproblem_W}
    \begin{align}
        \min_{\mathbf{W}} \quad &\sum_{k=1}^K \|\mathbf{w}_k\|^2 \\
        \mathrm{s.t.} \quad & \frac{1}{\gamma_k}\left| \mathbf{u}_k^H \mathbf{w}_k \right|^2 \ge \sum_{i=1, i \neq k}^K  \left| \mathbf{u}_k^H \mathbf{w}_i \right|^2 + \sigma_k^2, \forall k.
    \end{align}
\end{subequations}
Note that the above problem is the conventional power minimization problem for multi-user transmit beamforming. This problem can be transformed into an equivalent convex second-order cone (SOC) programming, and then solved by the fixed-point iteration. After obtaining the optimal $\mathbf{W}$, the optimal $\mathbf{V}$ can be calculated as 
\begin{equation}
    \mathbf{V} = \frac{\mathbf{W}}{\|\mathbf{W}\|_F}.
\end{equation}  

\subsubsection{Subproblem With Respect to $\mathbf{U}$} This subproblem can be expressed as 
\begin{subequations} \label{subproblem_U}
    \begin{align}
        \min_{\mathbf{U}, P > 0} \quad &P + \frac{1}{\rho} \| \mathbf{U} - \sum_{m=1}^M \mathbf{U}_m \|_F^2 \\
        \mathrm{s.t.} \quad & \frac{1}{\gamma_k}\left| \mathbf{u}_k^H \mathbf{v}_k \right|^2 \ge \sum_{i=1, i \neq k}^K  \left| \mathbf{u}_k^H \mathbf{v}_i \right|^2 + \frac{\sigma_k^2}{P}, \forall k.
    \end{align}
\end{subequations}
This subproblem is non-convex due to the quadratic term $\left| \mathbf{u}_k^H \mathbf{v}_k \right|^2$ on the left-hand side of the inequality constraints. We exploit the successive convex approximation method to address it. Let $\mathbf{U}^t = \big[\mathbf{u}_1^t,\dots,\mathbf{u}_k^t\big]$ denote the solution of $\mathbf{U}$ obtained in the previous iteration of the inner loop. Then, according to the first-order Taylor expansion, a low bound on the quadratic term $\left| \mathbf{u}_k^H \mathbf{v}_k \right|^2$ can be obtained as follows:
\begin{align}
    \left| \mathbf{u}_k^H \mathbf{v}_k \right|^2 &\ge - \left( \mathbf{u}_k^t\right)^H \mathbf{v}_k \mathbf{v}_k^H \mathbf{u}_k^t \nonumber \\
    &+ 2 \Re \left\{ \left( \mathbf{u}_k^t\right)^H \mathbf{v}_k \mathbf{v}_k^H \mathbf{u}_k \right\} \triangleq h \left( \mathbf{u}_k | \mathbf{u}_k^t \right),
\end{align} 
where the equality is achieved at $\mathbf{u}_k = \mathbf{u}_k^t$. Therefore, problem \eqref{subproblem_U} can be approximated by 
\begin{subequations} \label{subproblem_U_sca}
    \begin{align}
        \min_{\mathbf{U}, P > 0} \quad & P + \frac{1}{\rho} \|\mathbf{U} - \sum_{m=1}^M \mathbf{U}_m \|_F^2  \\
        \mathrm{s.t.} \quad & \frac{1}{\gamma_k} h \left( \mathbf{u}_k | \mathbf{u}_k^t \right) \ge \sum_{i = 1, i \neq k}^K \left| \mathbf{u}_k^H \mathbf{v}_i \right|^2 + \frac{\sigma_k^2}{P}, \forall k.
    \end{align}
\end{subequations}
This problem is a convex and therefore can be solved optimally by standard convex program solvers such as CVX.

\subsubsection{Subproblem With Respect to $\mathbf{U}_m$}
This subproblem is given by 
\begin{align}
    \min_{\mathbf{U}_m} \quad &\| \mathbf{U} - \sum_{m=1}^M \mathbf{U}_m \|_F^2 + \sum_{m=1}^M \| \mathbf{U}_m - \mathbf{\Phi}_m(\mathbf{X}) \|_F^2.
\end{align}
This problem is an unconstrained convex optimization problem, allowing us to use the first-order optimality condition to determine the optimal solution. By computing the partial derivatives of the objective function with respect to each $\mathbf{U}_m$ and setting these derivatives to zero, we derive the following conditions for the optimal solution:
\begin{equation}
    \sum_{i=1}^M \mathbf{U}_i + \mathbf{U}_m = \mathbf{U} + \mathbf{\Phi}_m(\mathbf{X}), \forall m,
\end{equation}
while yields the following optimal solution:
\begin{align} \label{subproblem_U_optimal}
    \mathbf{U}_m = \frac{1}{M+1} \left( \mathbf{U} + M \mathbf{\Phi}_m(\mathbf{X}) - \sum_{i=1, i \neq m}^M \mathbf{\Phi}_i(\mathbf{X})    \right).
\end{align}
\subsubsection{Subproblem With Respect to $\mathbf{X}$}
This subproblem is given by 
\begin{subequations} \label{subproblem_theta}
    \begin{align}
        \min_{\mathbf{X}} \quad & \sum_{m=1}^M \| \mathbf{U}_m - \mathbf{\Phi}_m(\mathbf{X})\|_F^2 \\
        \label{sub_theta_constraint}
        \mathrm{s.t.} \quad &  x_{n,m} - x_{n,m-1} \ge \Delta x, \forall n, m\neq 1, \\
        & x_{n,m} \in \mathcal{S}.
    \end{align}
\end{subequations}
The objective function of this problem can be reformulated as 
\begin{equation}
    \sum_{m=1}^M \| \mathbf{U}_m - \mathbf{\Phi}_m(\mathbf{X}) \|_F^2 = \sum_{n=1}^N \sum_{m=1}^M f(x_{n,m}),
\end{equation}
where 
\begin{align}
    f(x_{n,m}) = \sum_{k=1}^K \left| u_{k,n,m}  - \frac{\eta \alpha_m }{r_{k,n,m}} e^{j\left( \beta_0 r_{k,n,m} + \beta_{\mathrm{g}} x_{n,m} \right)} \right|^2,
\end{align}
with $u_{k,n,m}$ being the entry in the $n$-th row and $k$-th column of $\mathbf{U}_m$. It can be observed that each optimization variable $x_{n,m}$ is decoupled in the objective function. However, these variables remain coupled in the constraint \eqref{sub_theta_constraint}. 

\begin{algorithm}[tb]
    \caption{Element-wise Algorithm for Solving \eqref{subproblem_theta}}
    \label{alg:element_wise}
    \begin{algorithmic}[1]
        \STATE{initialize the optimization variables}
        \REPEAT
            \FOR{$n \in  \{1,\dots,N\}$ and $m \in \{ 1,\dots,M \}$}
            \STATE{update $x_{n,m}$ by solving problem \eqref{search_problem_1} through one-dimensional search}
            \ENDFOR
        \UNTIL{the fractional decrease of the objective value of problem \eqref{subproblem_theta} falls below a predefined threshold}
    \end{algorithmic}
\end{algorithm}

To tackle this issue, we propose an element-wise alternating optimization algorithm to solve problem \eqref{subproblem_theta}, where each $x_{n,m}$ is optimized by fixed the others. Specifically, the subproblem with respect to $x_{n,m}$ is given by 
\begin{subequations} \label{search_problem_1}
    \begin{align}
        \min_{x_{n,m}} \quad & f(x_{n,m}) \\
        \label{search_problem_1_constraint}
        \mathrm{s.t.} \quad & x_{n,m} \in \mathcal{S}_{n,m},
    \end{align}
\end{subequations}
where 
\begin{equation} \label{search_set}
    \mathcal{S}_{n,m} = [x_{n,m-1}+\Delta x, x_{n,m+1} - \Delta x] \cap \mathcal{S}.
\end{equation}
\textcolor{black}{It can be observed that problem \eqref{search_problem_1} is to find the minimum of a single-variable function over a finite interval. Depending on whether the activation is continuous or discrete, the solution can be obtained as follows:
\begin{itemize}
    \item For \emph{continuous activation}, to solve the multimodal objective function, an efficient approach is one-dimensional grid search \cite{10144718}, which prevents the optimization from getting stuck in local optima. In particular, the feasible set $\mathcal{S} = [0, x_{\max}]$ is discretized into a fine grid. The optimal solution that minimizes $g(x_{n,m})$ is then searched over the fine grid subject to the constraint~\eqref{search_problem_1_constraint}. To mitigate off-grid errors, a gradient-based refinement (e.g., Newton's method) can be applied starting from the grid-search solution, which, however, additional higher computational complexity.
    \item For \emph{discrete activation}, the optimal solution can be directly identified by searching over the discrete feasible set $\mathcal{S} = \{0, \frac{x_{\max}}{Q-1}, \dots,x_{\max} \}$, subject to the constraint~\eqref{search_problem_1_constraint}.
\end{itemize}
}
\noindent Based on the above discussion, both continuous and discrete positions of pinching antenna can be optimized through a one-dimensional search. Therefore, the overall element-wise algorithm for solving problem \eqref{subproblem_theta} is summarized in \textbf{Algorithm~\ref{alg:element_wise}}.

\subsubsection{Overall Algorithm}
Based on the above solutions for each subset of the optimization variables, the overall penalty-based alternating optimization algorithms is summarized in \textbf{Algorithm \ref{alg:PDD}}. \textcolor{black}{Specifically, $0< \epsilon < 1$ is used to reduce the penalty factor in the outer loop, and $\varepsilon$ is the constraint violation function, which is defined as the maximum value within the penalty terms:
\begin{equation} \label{constraint_violation}
    \varepsilon = \max \bigg\{ \| \mathbf{U} - \sum_{m=1}^M \mathbf{U}_m \|_{\infty}, \|\mathbf{U}_m -  \mathbf{\Phi}_m(\mathbf{X}) \|_{\infty}, \forall m \bigg\}.
\end{equation}}

\begin{algorithm}[tb]
    \caption{Penalty-based Alternating Optimization Algorithm for Joint Beamforming Problem \eqref{hybrid_beamforming_problem}}
    \label{alg:PDD}
    \begin{algorithmic}[1]
        \STATE{initialize the optimization variables, set $0 < \epsilon < 1$}
        \REPEAT
            \REPEAT
            \STATE{update $\mathbf{W}$ by solving problem \eqref{subproblem_W} and calculate $\mathbf{V} = \mathbf{W}/\|\mathbf{W}\|_F$  }
            \STATE{update $\mathbf{U}$ by solving problem \eqref{subproblem_U_sca}}
            \STATE{update $\mathbf{\mathbf{U}}_m$ as \eqref{subproblem_U_optimal}}
            \STATE{update $\mathbf{X}$ by \textbf{Algorithm \ref{alg:element_wise}} }
            \UNTIL{convergence}
            \STATE{update the penalty factor as $\rho \leftarrow \epsilon \rho$ }
        \UNTIL{$\varepsilon$ falls below a predefined threshold}
    \end{algorithmic}
\end{algorithm}

\noindent The computational complexity of each iteration in \textbf{Algorithm \ref{alg:PDD}} is determined by the complexity of updating its key variables. Specifically, updating $\mathbf{V}$ requires solving problem \eqref{subproblem_W}, which involves optimizing $NK$ variables with $K$ SOC constraints of dimension $NK$, resulting in a complexity of $O(N^4K^2)$. Similarly, updating $\mathbf{U}$ involves solving problem \eqref{subproblem_U_sca}, where $(NK+1)$ variables are optimized subject to $K$ SOC constraints of dimension $(NK+1)$, leading to a complexity of $O((NK+1)^4)$. For the update of $\mathbf{U}_m$ for all $m$, the computation of the closed-form expression in \eqref{subproblem_U_optimal} incurs a complexity of $O(MNK)$. Finally, updating $\mathbf{X}$ requires running \textbf{Algorithm \ref{alg:element_wise}}, which has a complexity of $O(I_{\mathrm{iter}} Q M N K)$, where $I_{\mathrm{iter}}$ denotes the number of iterations and $Q$ represents the size of the search space for the one-dimensional search. 

\textcolor{black}{Although \textbf{Algorithm~\ref{alg:PDD}} exhibits relatively high computational complexity, it possesses two important advantages. On the one hand, it directly addresses the original problem formulation, and the penalty-based method employed guarantees convergence to a stationary point~\cite{shi2020penalty, wu2020joint}. Consequently, \textbf{Algorithm~\ref{alg:PDD}} serves as a reliable baseline for evaluating other approaches, such as the zero-forcing algorithm presented in the subsequent section. On the other hand, \textbf{Algorithm~\ref{alg:PDD}} is highly generalizable, as it addresses coupling issues caused by pinching antennas by optimizing penalty terms independently of specific objective functions. Thus, it can be extended to address other optimization problems arising in pinching-antenna designs.}


\subsection{ZF-based Low-complexity Algorithm}

In this section, we further propose a low-complexity alternating optimization algorithm by using the ZF beamforming, which is capable of eliminate the inter-user interference completely. In this case, the SINR expression can be simplified significantly, thus reducing the optimization complexity. 

Given that $N \ge K$, we can obtain the following ZF beamforming matrix for any given pinching beamforming matrix \cite{bjornson2014optimal}:
\begin{equation} \label{ZF_BF}
    \mathbf{W} = \mathbf{\Psi}(\mathbf{X}) \left( \mathbf{\Psi}^H(\mathbf{X}) \mathbf{\Psi}(\mathbf{X}) \right)^{-1} \mathbf{P}^{\frac{1}{2}},
\end{equation}  
where $\mathbf{\Psi}(\mathbf{X}) \triangleq \mathbf{G}^H(\mathbf{X}) \mathbf{H}(\mathbf{X})$ is the equivalent channel matrix and $\mathbf{P} \in \mathbb{R}^{K \times K}$ is the diagonal power control matrix, given by 
\begin{equation}
    \mathbf{P} = \mathrm{diag}\left( P_1,\dots,P_K \right),
\end{equation}  
and $P_k$ is the power coefficient for user $k$. Consequently, the transmit power can be reformulated as 
\begin{align}
    \sum_{k=1}^K \|\mathbf{w}_k\|^2 = &\mathrm{tr}\left( \mathbf{W} \mathbf{W}^H \right) = \mathrm{tr}\left( \left( \mathbf{\Psi}^H(\mathbf{X}) \mathbf{\Psi}(\mathbf{X}) \right)^{-1} \mathbf{P} \right).
\end{align}
Substituting \eqref{ZF_BF} into \eqref{SINR_formula} yields the following simplified SINR expression:
\begin{equation}
    \mathrm{SINR}_k = \frac{P_k}{\sigma_k^2}.
\end{equation} 
Therefore, the transmit power minimization problem in \eqref{hybrid_beamforming_problem} can be simplified as 
\begin{subequations} \label{ZF_BF_problem}
    \begin{align}
        \min_{\mathbf{P}, \mathbf{X}} \quad & \mathrm{tr}\left( \left( \mathbf{\Psi}^H(\mathbf{X}) \mathbf{\Psi}(\mathbf{X}) \right)^{-1} \mathbf{P} \right) \\
        \mathrm{s.t.} \quad & \frac{P_k}{\sigma_k^2} \ge \gamma_k, \forall k, \\
        & \eqref{theta_constraint_1}, \eqref{theta_constraint_2}.
    \end{align}
\end{subequations}
It can be readily shown that the optimal $P_k, \forall k,$ to problem \eqref{ZF_BF_problem} is irrelevant to $\mathbf{X}$, which is given by 
\begin{equation} \label{optimal_ZF_P}
    P^{\mathrm{opt}}_k = \gamma_k \sigma_k^2, \forall k.
\end{equation}  
Therefore, the alternating optimization is no longer required. Given the optimal $\mathbf{P}^{\mathrm{opt}} = \mathrm{diag}(P^{\mathrm{opt}}_1,\dots,P^{\mathrm{opt}}_K )$, the optimization problem with respect to $\mathbf{X}$ is given by 
\begin{subequations} \label{ZF_BF_problem_Theta}
    \begin{align}
        \label{ZF_BF_problem_obj}
        \min_{\mathbf{X}} \quad & \mathrm{tr}\left( \left( \mathbf{\Psi}^H(\mathbf{X}) \mathbf{\Psi}(\mathbf{X}) \right)^{-1} \mathbf{P}^{\mathrm{opt}} \right) \\
        \mathrm{s.t.} \quad & \eqref{theta_constraint_1}, \eqref{theta_constraint_2}.
    \end{align}
\end{subequations}
This problem can also be solved using an element-wise one-dimensional search algorithm. Specifically, the optimal value of each element $x_{n,m}$ in $\mathbf{X}$ is determined by searching over the set $\mathcal{S}_{n,m}$ defined in \eqref{search_set}, by keeping all other entries fixed. 

However, searching for the optimal $x_{n,m}$ to minimize the objective function in \eqref{ZF_BF_problem_obj} can result in extremely high computational complexity, as matrix inversion must be computed at each search point.  To mitigate this issue, we propose a decomposition method for the objective function in \eqref{ZF_BF_problem_obj} in the case of $N > K$. Specifically, for optimizing $x_{n,m}, \forall m,$ associated with the $n$-th RF chain, we introduce the following decomposition:  
\begin{equation}
    \mathbf{\Psi}^H(\mathbf{X}) \mathbf{\Psi}(\mathbf{X}) = \mathbf{a}_n \mathbf{a}_n^H + \mathbf{B}_n \mathbf{B}_n^H, 
\end{equation}  
where $\mathbf{a}_n \in \mathbb{C}^{K \times 1}$ is the $n$-th column vector of $\mathbf{\Psi}^H(\mathbf{X})$, while $\mathbf{B}_n \in \mathbb{C}^{K \times (N-1)}$ is the matrix containing all column vectors of $\mathbf{\Psi}^H(\mathbf{X})$ except $\mathbf{a}_n$. According to \eqref{PH_formula}, the expression of $\mathbf{a}_n$ is given by 
\begin{equation}
    \mathbf{a}_n = \left[ \mathbf{h}_1^H(\mathbf{x}_n)\mathbf{g}(\mathbf{x}_n),\dots,\mathbf{h}_K^H(\mathbf{x}_n)\mathbf{g}(\mathbf{x}_n) \right]^T.
\end{equation} 
The expression of $\mathbf{B}_n$  can be derived in a similar manner. Given that $N > K$, the matrix $\mathbf{B}_n \mathbf{B}_n^H$ must be full-rank. As a result, the matrix inversion in the objective function \eqref{ZF_BF_problem_obj} can be rewritten as 
\begin{align} \label{ZF_decomposition}
    &\left( \mathbf{\Psi}^H(\mathbf{X}) \mathbf{\Psi}(\mathbf{X}) \right)^{-1} = \left( \mathbf{a}_n \mathbf{a}_n^H + \mathbf{B}_n \mathbf{B}_n^H \right)^{-1} \nonumber \\
    &\hspace{0.5cm} \overset{(a)}{=} \left(\mathbf{B}_n \mathbf{B}_n^H\right)^{-1} -  \frac{\left(\mathbf{B}_n \mathbf{B}_n^H\right)^{-1} \mathbf{a}_n \mathbf{a}_n^H \left(\mathbf{B}_n \mathbf{B}_n^H\right)^{-1}}{1 + \mathbf{a}_n^H \left(\mathbf{B}_n \mathbf{B}_n^H\right)^{-1} \mathbf{a}_n},
\end{align} 
where step $(a)$ follows from the Sherman-Morrison formula. By substituting \eqref{ZF_decomposition} into \eqref{ZF_BF_problem_Theta}, the subproblem for optimizing $x_{n,m}$ is formulated as
\begin{subequations} \label{ZF_search_problem}
    \begin{align}
        \max_{x_{n,m}} \quad & \frac{ \mathbf{a}_n^H \left(\mathbf{B}_n \mathbf{B}_n^H\right)^{-1} \mathbf{P}^{\mathrm{opt}} \left(\mathbf{B}_n \mathbf{B}_n^H\right)^{-1} \mathbf{a}_n}{1 + \mathbf{a}_n^H \left(\mathbf{B}_n \mathbf{B}_n^H\right)^{-1} \mathbf{a}_n} \\
        \mathrm{s.t.} \quad & x_{n,m} \in \mathcal{S}_{n,m}.
    \end{align}
\end{subequations}
When applying the one-dimensional search method to solve this problem, only a single matrix inversion, $\left(\mathbf{B}_n \mathbf{B}_n^H\right)^{-1}$, needs to be computed for all search points, which significantly reduce the computational complexity.  

\begin{algorithm}[tb]
    \caption{ZF-based Low-complexity Algorithm for Joint Beamforming Problem \eqref{hybrid_beamforming_problem}}
    \label{alg:ZF}
    \begin{algorithmic}[1]
        \STATE{calculate $P_k, \forall k,$ as \eqref{optimal_ZF_P}, and initialize $\mathbf{X}$}
        \REPEAT
            \FOR{$n \in  \{1,\dots,N\}$ and $m \in \{ 1,\dots,M \}$}
            \STATE{update $x_{n,m}$ by solving problem \eqref{ZF_search_problem} through one-dimensional search}
            \ENDFOR
        \UNTIL{the fractional decrease of the objective value of problem \eqref{ZF_BF_problem_obj} falls below a predefined threshold}
        \STATE{calculate $\mathbf{W}$ as \eqref{ZF_BF}}
    \end{algorithmic}
\end{algorithm}

The overall ZF-based low-complexity algorithm is summarized in \textbf{Algorithm \ref{alg:ZF}}. The computational complexity of each iteration is analyzed as follows. In particular, for each $n$, the matrix inversion $\left(\mathbf{B}_n \mathbf{B}_n^H\right)^{-1}$ need to be computed, resulting in a complexity of a complexity of $O(K^3)$. Furthermore, The one-dimensional search has a complexity of $O(Q K^2)$. Therefore, the overall computational complexity per iteration is given by $O(N K^2 + Q MN K^2)$.

\section{Numerical Results} \label{sec:results}

\begin{figure}[t!]
  \centering
  \includegraphics[width=0.48\textwidth]{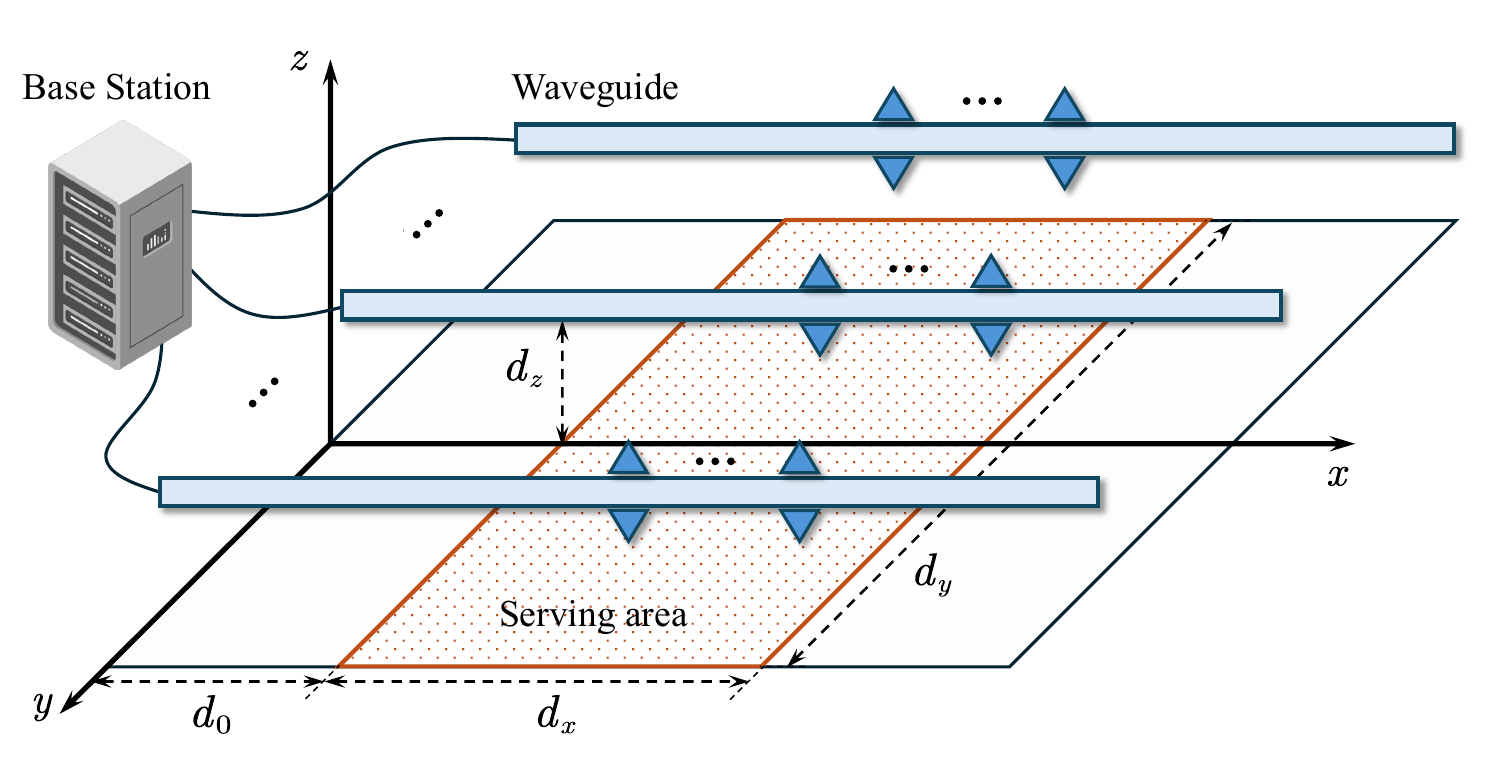}
  \caption{The simulation setup.}
  \label{setup}
\end{figure}

In this section, numerical results are presented to illustrate the advantages of PASS and validate the effectiveness of the proposed algorithms. \textcolor{black}{Unless stated otherwise, we consider an indoor communication scenario, and the following setup is used throughout the simulations.} We consider $N = 5$ waveguides, each equipped with $M = 6$ pinching antennas and fed by a dedicated RF chain, serving $K = 4$ users. The geometric configuration of the setup is depicted in Fig. \ref{setup}, where the key parameters are set as $d_0 = 15$ m, $d_x = 30$ m, $d_y = 3$ m, and $d_z = 10$ m. The waveguides are uniformly deployed in parallel along the $x$-axis with a spacing of $6$ m, while the users are randomly positioned within the designated service area. The maximum range and minimum spacing of pinching antenna positions are set to $x_{\max} = 50$ m and $\Delta x = 0.1$ m, respectively. For discrete activation, the number of discrete positions is set to $10$ per meter. 

Furthermore, the carrier frequency, channel gain, noise power, and effective index of the waveguide are set to $15$ GHz, $|\eta|^2 = \left(\frac{\lambda}{4 \pi}\right)^2 = -56$ dB, $\sigma_k^2 = -80$ dBm, and $n_g = 1.4$, respectively. The total power radiated from all pinching antennas along each waveguide is constrained by $\sum_{m=1}^{M} \alpha_m^2 = 0.9$, which applies to both equal and proportional power models. The minimum SINR of all users is set to $\gamma_k = 20$ dB. For the proposed algorithms, the penalty factor is initialized as $\rho = 10$, the reduction factor is set to $\epsilon = 0.1$, and the convergence threshold is set to $10^{-3}$. The number of search points in the one dimensional search for continuous activation is set to $10^6$. Furthermore, due to its sensitivity to initialization, the penalty-based method utilizes the antenna positions obtained from the ZF-based algorithm as its starting point. Unless otherwise specified, all subsequent results are obtained by averaging over $100$ random samples of user positions.

\begin{figure*}[t!]
  \centering
  \begin{subfigure}[t]{0.32\textwidth}
    \includegraphics[width=1\textwidth]{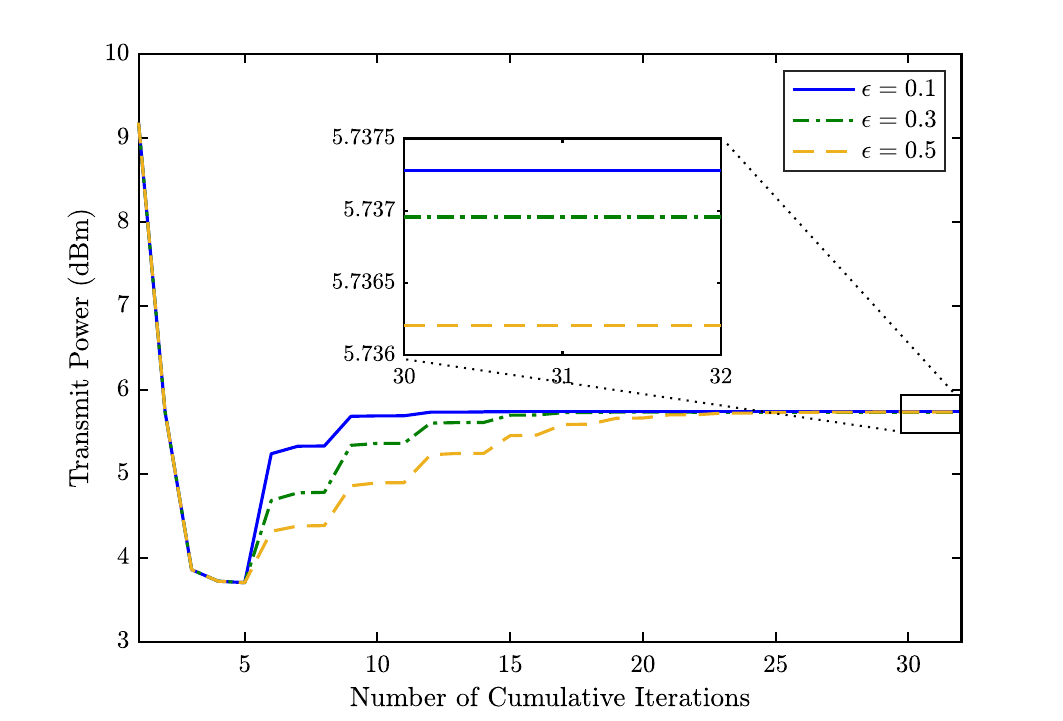}
    \caption{\textcolor{black}{Transmit power in \textbf{Algorithm \ref{alg:PDD}}.}}
    \label{convergence_1}
  \end{subfigure}
  \begin{subfigure}[t]{0.32\textwidth}
    \includegraphics[width=1\textwidth]{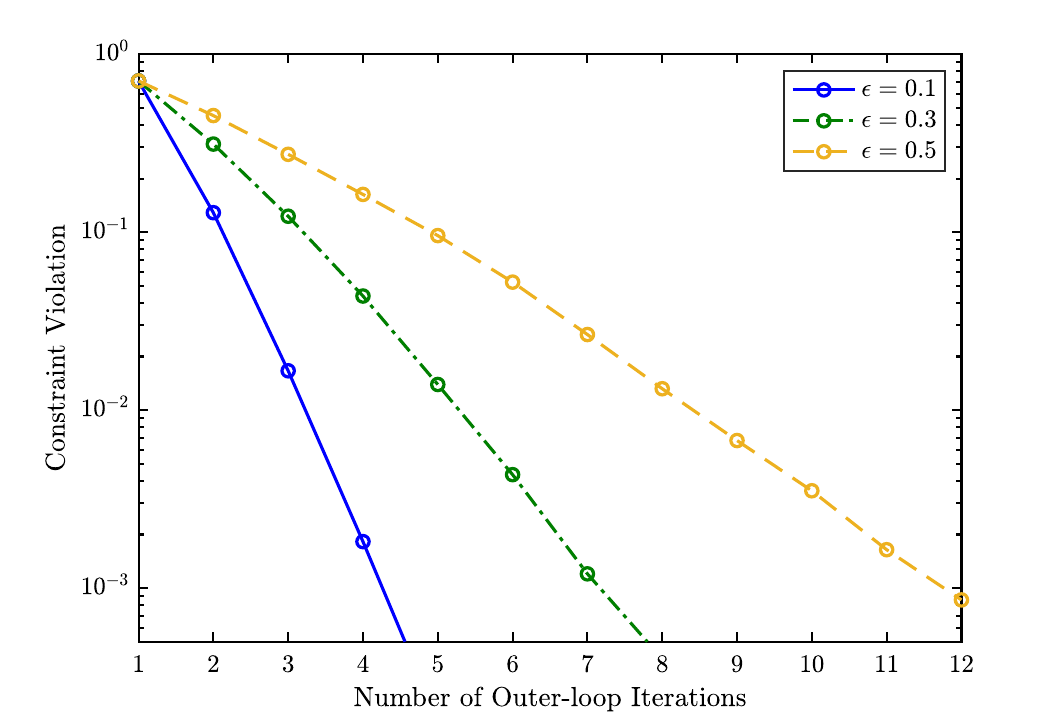}
    \caption{\textcolor{black}{Constraint violation \eqref{constraint_violation} in \textbf{Algorithm \ref{alg:PDD}}.}}
    \label{convergence_2}
  \end{subfigure}
  \begin{subfigure}[t]{0.32\textwidth}
    \includegraphics[width=1\textwidth]{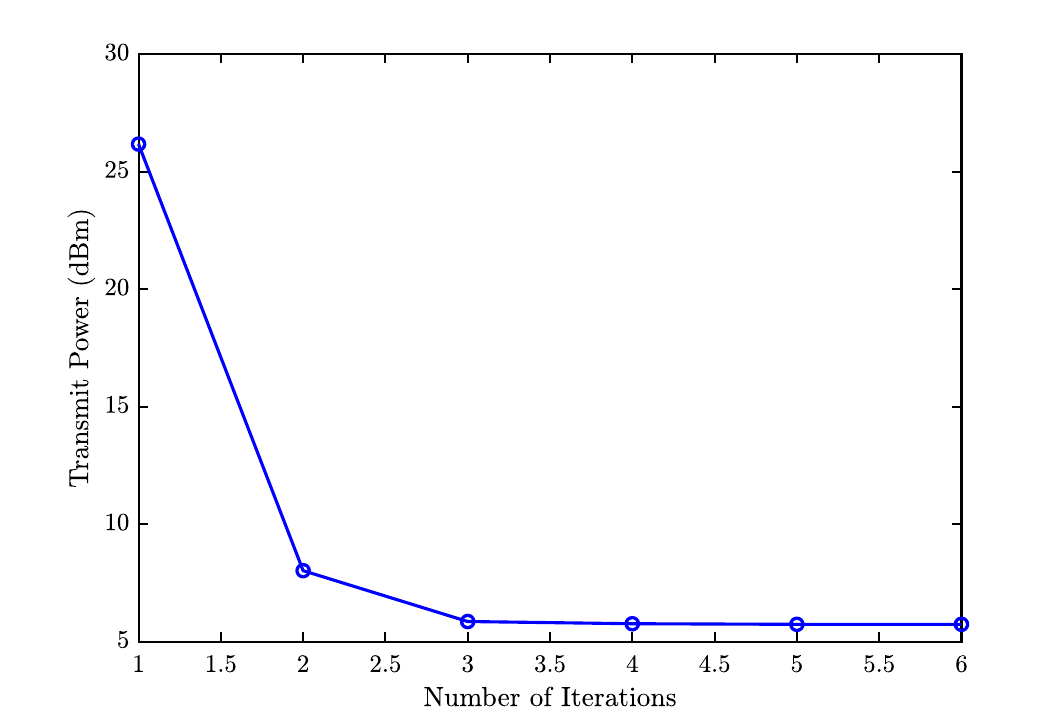}
    \caption{Transmit power in \textbf{Algorithm \ref{alg:ZF}}.}
    \label{convergence_3}
  \end{subfigure}
  \caption{Convergence behavior of the proposed algorithms.}
  \label{convergence}
  \vspace{-0.5cm}
\end{figure*} 

For performance comparison, we consider the following benchmark schemes:
\begin{itemize}
  \item \textbf{Conventional MIMO:} In this benchmark, a conventional MIMO BS is positioned at $(0,0,3)$ and equipped with a uniform linear array along $x$-axis  comprising $N_{\mathrm{RF}} = 5$ chains, each connected to a single antenna. The antenna spacing is set to half the wavelength\footnote{\textcolor{black}{Movable antennas have recently garnered significant attention. Their primary advantage lies in enhancing spatial DoFs when only a limited number of antennas are deployed within a given aperture. However, in our baselines, we consider a dense deployment with half-wavelength spacing, which is known to fully exploit the available spatial DoFs \cite{11005662}. Therefore, we assume the antennas to be fixed and non-movable in our baselines.}}. Under this setup, the signal received by user $k$ is given by 
  \begin{equation}
    \overline{y}_k = \overline{\mathbf{h}}_k^H \sum_{i=1}^K \mathbf{w}_i c_i + n_k,
  \end{equation}
  where $\overline{\mathbf{h}}_k \in \mathbb{C}^{N \times 1}$ is the channel vector. The $k$-th entry of $\overline{\mathbf{h}}_k$ is $\frac{\eta}{\overline{r}_{n,k}}e^{j \beta_0 \overline{r}_{n,k}}$, with $\overline{r}_{n,k}$ denoting the distance between $n$-th antenna and user $k$. The corresponding transmit power minimization problem can be solved using the method described in \cite{bjornson2014optimal}.
  
  \item \textbf{Massive MIMO:} In this benchmark, we consider a massive MIMO BS positioned at $(0,0,3)$, equipped with a uniform linear array with half-wavelength spacing along $x$-axis. The number of antennas and RF chains is set to $N = 30$ and $N_{\mathrm{RF}} = 5$, respectively, matching the configuration of the considered PASS system. Furthermore, since each RF chain in the PASS system is connected to only a subset of overall pinching antennas, we assume a similar sub-connected hybrid beamforming architecture at the massive MIMO BS, where each RF chain is connected to a subset of antennas via phase shifters \cite{yu2016alternating}. Under this setup, the signal received by user $k$ is given by 
  \begin{equation}
    \overline{\overline{y}}_k = \overline{\overline{\mathbf{h}}}_k^H \mathbf{W}_{\mathrm{RF}} \sum_{i=1}^K \mathbf{w}_i c_i + n_k,
  \end{equation}
  where $\overline{\overline{\mathbf{h}}}_k \in \mathbb{C}^{N \times 1}$ is the channel vector exhibiting the same form as $\overline{\mathbf{h}}_k$, and $\mathbf{W}_{\mathrm{RF}} \in \mathbb{C}^{N \times N_{\mathrm{RF}}}$ is the analog beamforming matrix realized by the phase shifters. The corresponding transmit power minimization problem is solved by integrating the algorithms in \cite{yu2016alternating} and \cite{shi2018spectral}.  
\end{itemize}

\subsection{Performance of the Proposed Algorithms}

\begin{figure}[t!]
  \centering
  \includegraphics[width=0.48\textwidth]{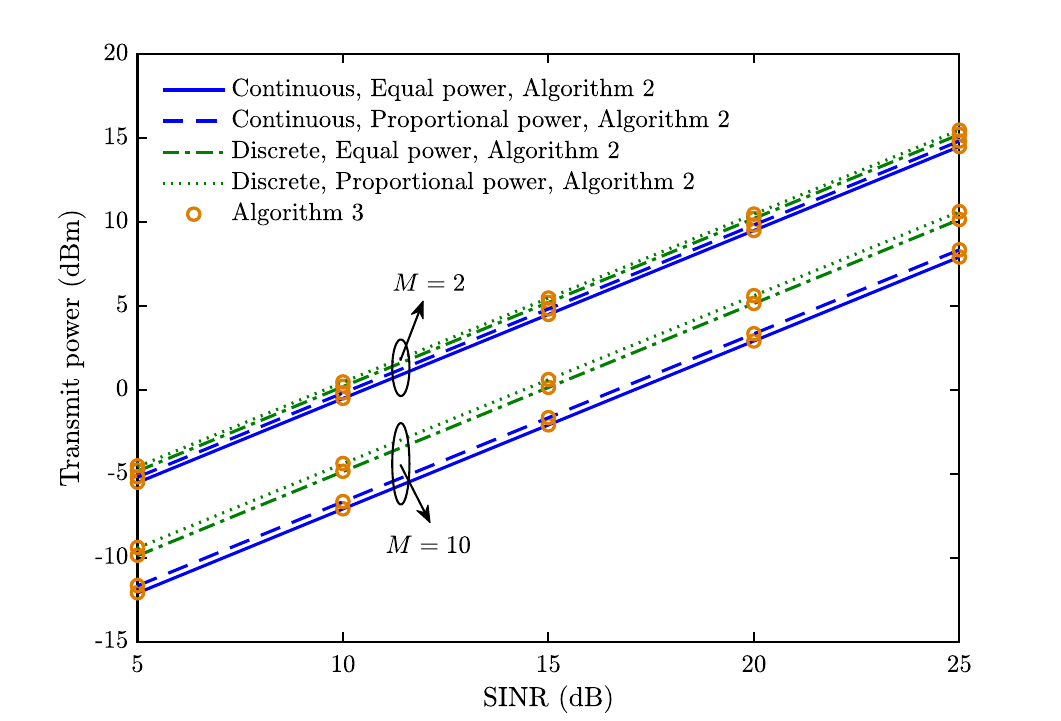}
  \caption{\textcolor{black}{Comparison between the proposed algorithms under different simulation setup.}}
  \label{power_algorithm_compare}
\end{figure}

Fig. \ref{convergence} illustrates the convergence behavior the proposed algorithms.  Specifically, the convergence behavior of the transmit power $P$ in \textbf{Algorithm \ref{alg:PDD}} is shown in Fig. \ref{convergence_1}. It is interesting to observe that the transmit power $P$ is not reduced monotonically as the iteration progresses, but exhibits an oscillating upward trend before converging to a stable value. This phenomenon is due to the utilization of the penalty method. Specifically, in the initial iterations, when the penalty factor $\rho$ is large, the transmit power $P$ is minimized with an almost unconstrained auxiliary channel matrix $\mathbf{U}$, allowing it to converge to a very low stable value in the inner loop. However, as the penalty factor decreases in the outer loop, constraint violations gradually diminish as shown in Fig. \ref{convergence_2}, forcing the auxiliary channel matrix $\mathbf{U}$ to conform to the channel structure and constraints in PASS. Consequently, the transmit power increases as the outer loop progresses. Fig. \ref{convergence_3} demonstrates the fast convergence of the proposed \textbf{Algorithm \ref{alg:ZF}}. Despite its low complexity, \textbf{Algorithm \ref{alg:ZF}} achieves performance comparable to \textbf{Algorithm \ref{alg:PDD}} across various system setups, as shown in Fig.~\ref{power_algorithm_compare}. 

Furthermore, Figs. \ref{convergence_1} and \ref{convergence_2} also reveal that different values of the reduction factor $\epsilon$ affect the convergence speed. \textcolor{black}{Specifically, the algorithm converges more quickly with a smaller $\epsilon$, while a larger $\epsilon$ leads to improved performance by allowing the algorithm to more thoroughly minimize transmit power before reducing the penalty terms.} However, it is worth noting that the performance gain from increasing $\epsilon$ from $0.1$ to $0.5$ is negligible. \textcolor{black}{In the remaining simulations, only the results obtained by \textbf{Algorithm \ref{alg:PDD}} are presented to focus on the performance gains achieved by PASS.}

\subsection{Impact of the Minimum SINR}

\begin{figure}[t!]
  \centering
  \includegraphics[width=0.48\textwidth]{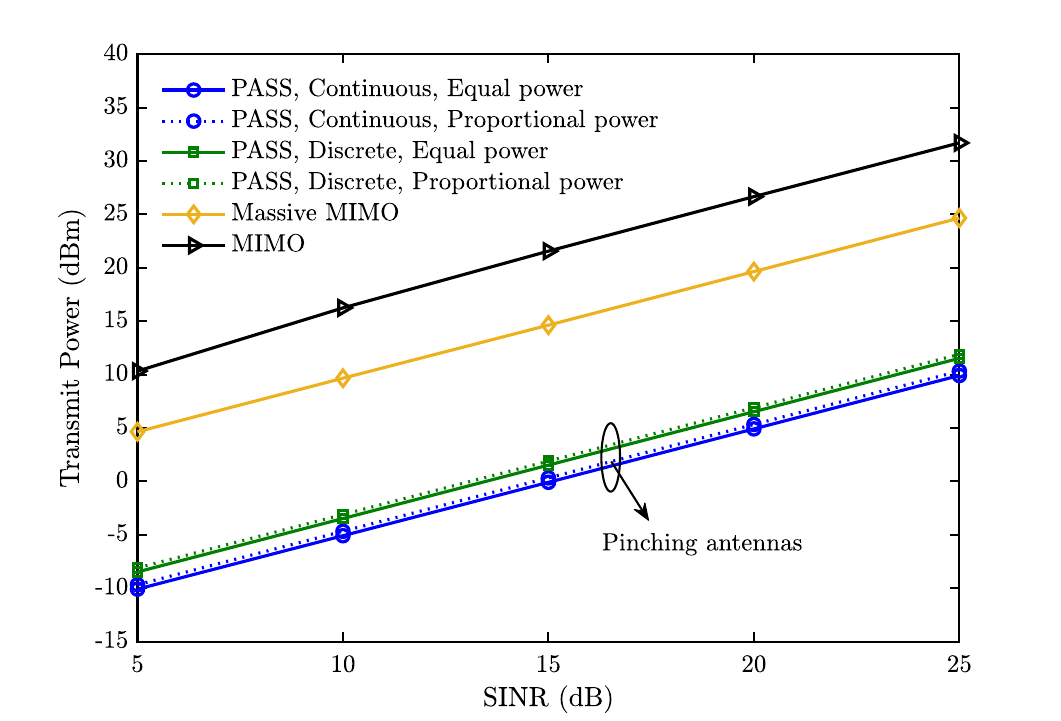}
  \caption{Transmit power versus the minimum SINR.}
  \label{power_vs_SINR}
\end{figure}

Fig. \ref{power_vs_SINR} shows the impact of the minimum SINR on transmit power. As expected, the transmit power increases with higher minimum SINR requirements, while PASS consistently achieves the lowest transmit power within the considered range. For instance, considering the PASS system with continuous activation and the equal power model, when the minimum SINR of users is 20 dB, PASS significantly reduces the transmit power by $99.3\%$, i.e., from $26.6$ dBm to $4.9$ dBm, compared to conventional MIMO, and by $96.6\%$, i.e., from $19.6$ dBm to $4.9$ dBm, compared to massive MIMO. This remarkable power reduction is primarily due to the ability of PASS to reduce free-space path loss significantly. It is also noteworthy that such an enhancement is achieved by incorporating only low-cost pinching antennas, rather than relying on massive expensive phase shifters as in massive MIMO.

It can also be observed from Fig. \ref{power_vs_SINR} that the proportional power model exhibits almost an identical performance as the equal power model, unveiling the negligible impact of unbalanced antenna efficiency in PASS. Furthermore, while discretely deploying pinching antennas results in non-negligible performance loss, the system still achieves a significant reduction in transmit power compared to both conventional MIMO and massive MIMO. For example, when the minimum SINR is $20$ dB, the PASS with discrete activation reduce the transmit power by $95\%$ and $99\%$ compared to conventional MIMO and massive MIMO, respectively.

\subsection{Impact of the Distance}

\begin{figure}[t!]
  \centering
  \includegraphics[width=0.48\textwidth]{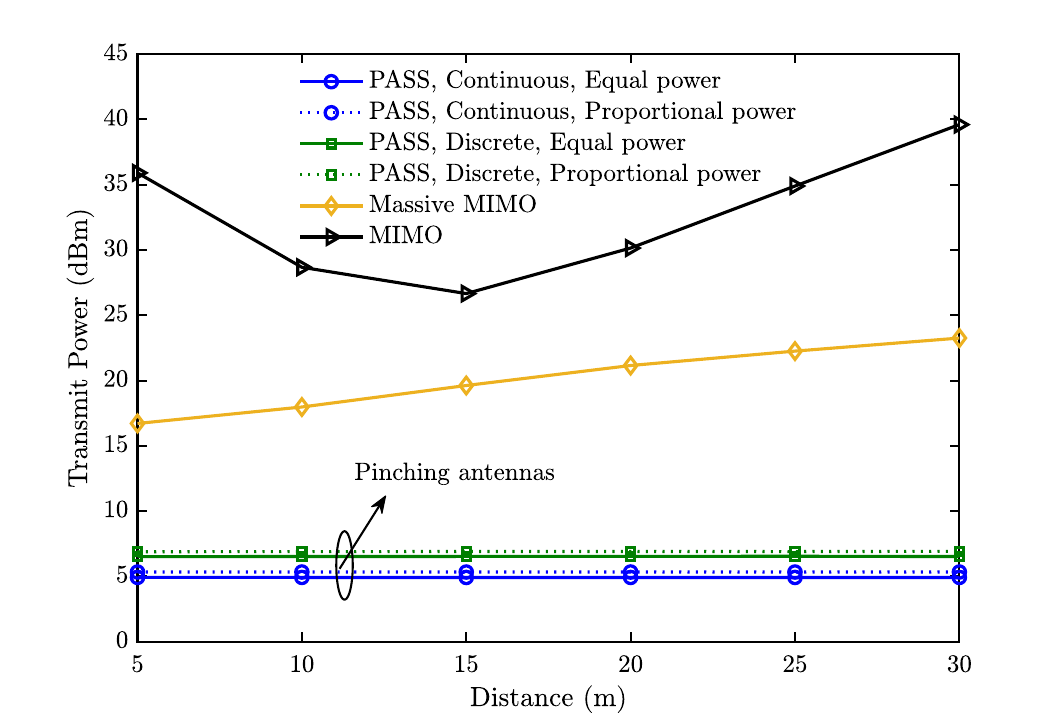}
  \caption{Transmit power versus the distance $d_0$.}
  \label{power_vs_distance}

  \includegraphics[width=0.48\textwidth]{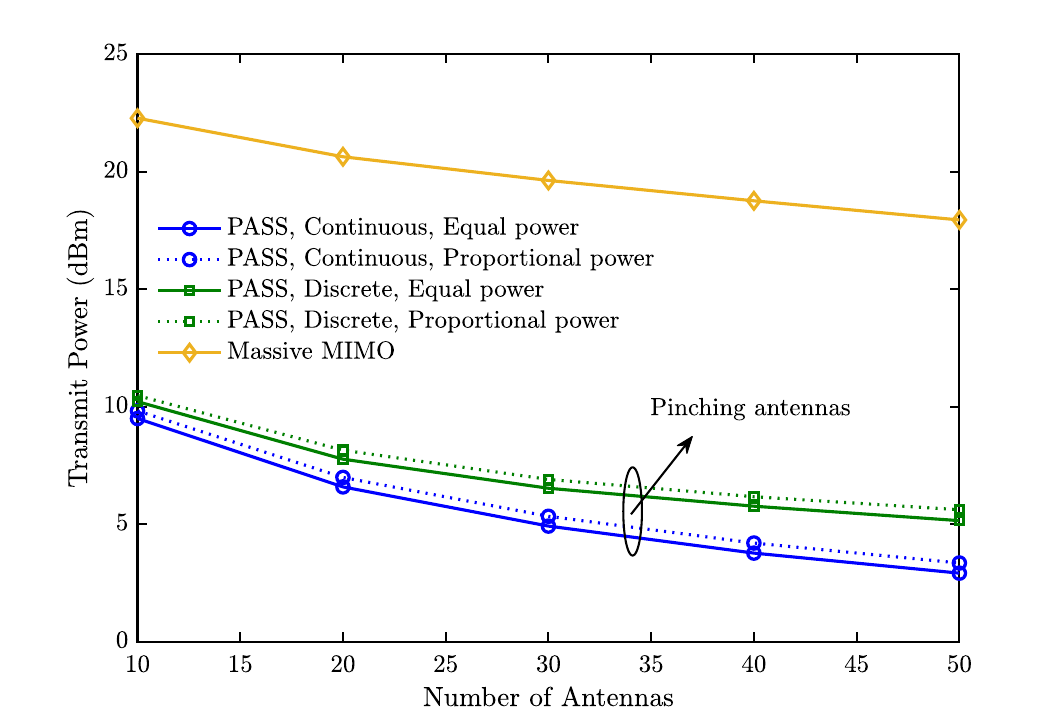}
  \caption{Transmit power versus the number of antennas.}
  \label{power_vs_antenna}
\end{figure}

Fig. \ref{power_vs_distance} examines the impact of distance $d_0$ on transmit power, where a larger $d_0$ indicates an increased separation between the BS and the serving area. It can be observed that the transmit power achieved by PASS remains almost unchanged as $d_0$ increases. This phenomenon can be attributed to two key factors. First, the pathloss associated with in-waveguide propagation is negligible. Second, the pinching antennas can always be deployed in proximity to the serving area, ensuring an almost constant free-space pathloss. 

Furthermore, it is somewhat counterintuitive that the transmit power in conventional MIMO initially decreases before increasing. This phenomenon arises because conventional MIMO performance is influenced not only by free-space path loss but also by the directivity of the antenna array. More specifically, the uniform linear array used in conventional MIMO can generate a highly directional beam when transmitting in its broadside direction (i.e., with a small angle of departure) \cite{kallnichev2001analysis}. However, as the transmission direction shifts closer to the array's end-fire (i.e., with a large angle of departure), its beamforming capability gradually weakens \cite{kallnichev2001analysis}, resulting in increased inter-user interference. In the considered simulation setup, as illustrated in Fig. \ref{setup}, when the distance $d_0$ is smaller, users are more likely to be positioned in the end-fire directions of the antenna array. This leads to significant inter-user interference, which, in turn, increases the required transmit power. This explains why, for conventional MIMO, its transmit power increases despite the reduction in free-space path loss as the serving area moves closer to the BS. However, for massive MIMO, a smaller distance always results in lower transmit power. This is because a large antenna array can significantly enhance beamforming capability even in the end-fire direction, mitigating the directivity limitations faced by conventional MIMO. Additionally, thanks to the flexibility of pinching antenna deployment, PASS can effectively address this directivity issue as well.

\subsection{Impact of the Number of Antennas}

Fig. \ref{power_vs_antenna} studies the impact of the number of antennas on transmit power. It can be observed that increasing the number of antennas leads to a reduction in transmit power for all the considered schemes. For instance, when the number of antennas increases from 10 to 50, the transmit power is reduced by $78\%$ for PASS with continuous activation and $68.8\%$ for PASS with discrete activation. This improvement is primarily due to the enhanced beamforming capability and the higher array gain provided by a larger antenna array, which not only mitigates inter-user interference but also strengthens the desired signal, resulting in more efficient power usage. 

\begin{figure}[t!]
  \centering
  \includegraphics[width=0.48\textwidth]{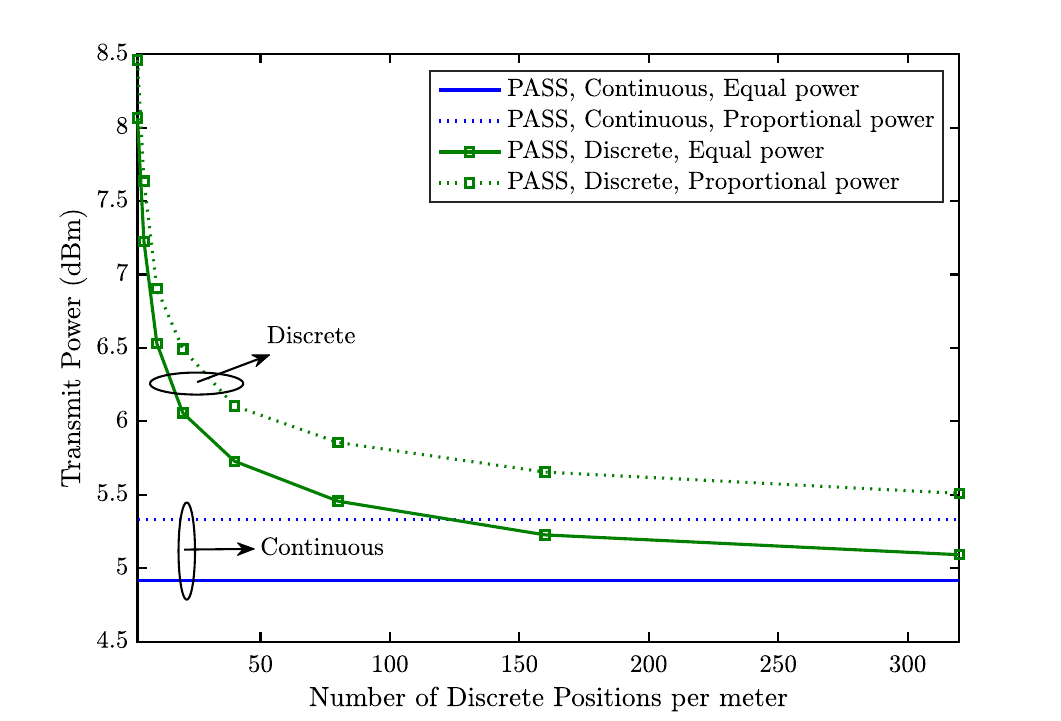}
  \caption{Transmit power versus the number of discrete positions.}
  \label{power_vs_discrete}

  \includegraphics[width=0.48\textwidth]{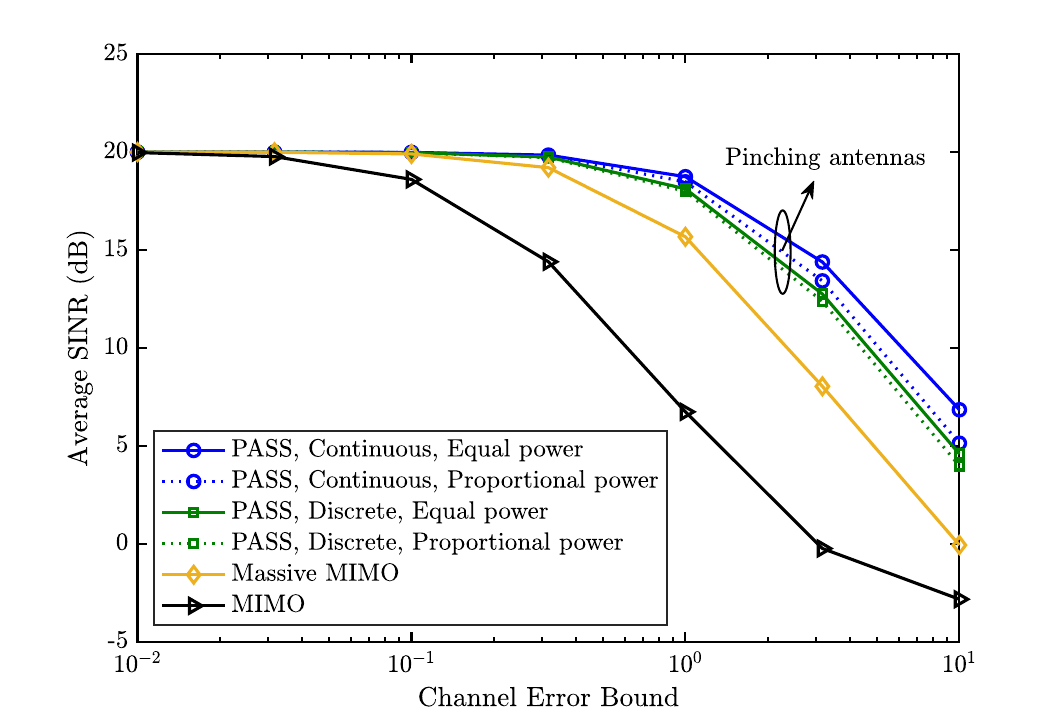}
  \caption{\textcolor{black}{Average SINR versus the channel error bound $\varepsilon_{\mathrm{est}}$.}}
  \label{sinr_vs_error}
\end{figure}


\subsection{Impact of the Number of Discrete Positions}

Fig. \ref{power_vs_discrete} provides more insights into the discrete activation of PASS, specifically focusing on the impact of the number of available discrete positions. As expected, the performance of discrete activation gradually approaches that of continuous activation as the number of discrete positions increases. However, achieving performance comparable to continuous activation requires a significantly large number of discrete positions, i.e., larger than $300$ per meter. This is because effective beamforming gain becomes challenging to achieve with a limited number of discrete positions. To understand this, consider the phase shift $e^{-j \beta_{\mathrm{g}} x_{\mathrm{p}}}$ induced by a pinching antenna located at position $x_{\mathrm{p}}$, where $\beta_{\mathrm{g}} = \frac{2 \pi n_{\mathrm{g}}}{\lambda} \approx 440$ under the given simulation setup. For optimal beamforming capability, $\beta_{\mathrm{g}} x_{\mathrm{p}}$ must be adjustable across the full range of $[0, 2\pi]$. However, due to the large value of $\beta_{\mathrm{g}}$, achieving this flexibility requires sampling $x_{\mathrm{p}}$ at very fine intervals, necessitating a high density of discrete positions. This result underscores the importance of developing high-resolution pinching antenna activation structures for the practical PASS implementation.    

\subsection{\textcolor{black}{Impact of the Channel Estimation Error}}

\textcolor{black}{
Fig. \ref{sinr_vs_error} explores the impact of the channel estimation error on the average SINR among users. We adopt a deterministic model to characterize the channel estimation error, which models the free-space channel as \cite{4838902} 
\begin{align} \label{channel_error}
  \mathbf{h}_k = \hat{\mathbf{h}}_k + \Delta \mathbf{h}_k, \quad \| \Delta \mathbf{h}_k \| \le \varepsilon_{\mathrm{est}}, \forall k,
\end{align}
where $\hat{\mathbf{h}}_k$ is the estimated channel for user $k$ and $\Delta \mathbf{h}_k$ denotes the channel estimation error with its norm bounded by $\varepsilon_{\mathrm{est}} > 0$. For PASS, the in-waveguide channel depends solely on the positions of the pinching antennas and thus can be calculated accurately with negligible error. Channel estimation errors for conventional MIMO and massive MIMO systems can be similarly modeled using \eqref{channel_error}. The results in Fig. \ref{sinr_vs_error} is obtained by optimizing the system with the estimated channel $\hat{\mathbf{h}}_k$ and evaluating the performance over the actual channel $\mathbf{h}_k$. As expected, increasing channel error bound $\varepsilon_{\mathrm{est}}$ degrades SINR, however, PASS consistently achieves the highest SINR, demonstrating its robustness to channel estimation errors.
}
\section{Conclusions} \label{sec:conclusion}

This paper has proposed a physics-based hardware model and several signal models for the emerging PASS technology, serving as a valuable reference for future research. Two efficient optimization algorithms were proposed to tackle the joint transmit and pinching beamforming optimization problem in downlink multi-user PASS. Numerical results demonstrated the significant advantages of PASS in reducing transmit power compared to conventional wireless systems.

However, several key challenges remain unresolved. For instance, the proposed physics model has been applied only to the downlink scenario, whereas modeling for uplink scenarios may be more complex and requires further investigation. Additionally, this study considered the simplest transmit beamforming architecture, where each waveguide is connected to a single RF chain. Exploring hybrid analog-digital beamforming architectures with pinching beamforming, which support fully connected configurations and a flexible number of RF chains, is another promising direction for future research. \textcolor{black}{Finally, practical indoor or even outdoor deployment is also essential to fully realize the benefits of PASS.}


\bibliographystyle{IEEEtran}
\bibliography{reference/mybib}

\end{document}